\documentclass[preprint]{revtex4}

\newcommand{\bea}{\begin{eqnarray}}
\newcommand{\eea}{\end{eqnarray}}

\newcommand{\be}{\begin{equation}}
\newcommand{\ee}{\end{equation}}
\newcommand{\lbl}{\label}

\begin{document}

\begin{titlepage}

\title{Field Theoretic Formulation of Kinetic theory:\\ I. Basic
Development}
\thispagestyle{empty}

\author{Shankar P. Das$^{2,1}$ and Gene F. Mazenko$^{1}$}

\affiliation{$^1$The James Franck Institute and the Department of
Physics,
The University of Chicago, Chicago, Illinois 60637, U.S.A.\\
$^2$School of Physical Sciences, Jawaharlal Nehru University, New
Delhi - 110067, India.}

\vspace*{1cm}

\begin{abstract}

We show how kinetic theory, the statistics of classical particles
obeying Newtonian dynamics, can be formulated as a field theory.
The field theory can be organized to produce a self-consistent
perturbation theory expansion in an effective interaction potential.  The
need for a self-consistent approach is suggested by our interest
in investigating ergodic-nonergodic transitions in dense fluids.
The formal structure we develop has been implemented in detail
for the simpler case of Smoluchowski dynamics.  One aspect of the approach
is the identification of a core problem spanned by the variables
$\rho$ the number density and $B$ a response density.  In this
paper we set up the perturbation theory expansion with explicit
development at zeroth and first order.  We also determine all of
the cumulants in the noninteracting limit among the core variables
$\rho$ and $B$.
\end{abstract}


\maketitle
\end{titlepage}

\section{Introduction}

There exists a well defined approach to the problem of classical
many-particle dynamics. Kinetic theory governs the kinetics of
particles obeying Newtonian dynamics.
It is one of the oldest disciplines in all of science.  From
the early work of Bernoulli\cite{bern} to the seminal work of
Boltzmann\cite{bolt}
and Maxwell\cite{maxw}, kinetic theory has been applied to dilute systems
out of equilibrium. More systematic modern methods have been developed
for systems fluctuating in equilibrium.  Based on work by Koopman\cite{koop}
and von Neumann\cite{vonn}, one can express the time evolution
in terms of Koopman's operator $e^{i{\cal L}t}$ where ${\cal L}$ is
 the Liouville operator.
For example one can develop in a rather straight forward way density expansions for
transport coefficients\cite{zwan,boul} and
memory functions\cite{FRKT}. Similarly one can develop expansions in
terms of the interaction potential\cite{akch,fors}.
However these approaches have their  limitations.
It is not known how to systematically rearrange the respective density or
potential expansions in a self-consistent manner.
By self-consistent we mean here the interaction kernels of the associated
kinetic equations can be expressed in terms of the full unknown
correlation functions.  Thus one obtains a nonlinear kinetic equation
that must be solved self-consistently.
While there are many formal reasons why this self-consistency is
desirable, our motivation for  pursuing such a theory in the present case
is more practical.  We want to understand the role of ergodic-nonergodic
(ENE) transitions\cite{close}. Self-consistency is essential if one is
to investigate
whether one does or does not have an ENE transition in dense fluids.

The liquid cooled to low temperature or compressed to
high densities reach a stage in which it behaves like a frozen
solid without any long range order. Understanding the formation
of the amorphous solid  state of the liquid coming from the ergodic
liquid state has remained an unresolved problem of physics.
A theory of the formation of the amorphous solid like state with self generated
disorder will require  developing techniques for testing  the possibilities of
ENE transitions in the dense liquid.
There is compelling evidence that as one approaches the glass transition
one comes close to
an ENE transition.  Even if we finally rule out the existence
of a physical ENE transition, we need a formal structure which
potentially can show such a transition.  We present
the outline of such a self-consistent theory here for the case of Newtonian
dynamics (ND).
Elsewhere one of us has developed\cite{FTSPD,SDENE}
such a self-consistent theory for a classical set of particles which obey the simpler
Smoulochowski dynamics\cite{SD}(SD). In that case a program of investigating
the existence of ENE transitions is well along and one has
a well defined approach to the problem. A key attribute of  this theory
is that well-defined approximations are available with well-defined
corrections. In this paper we show at the formal level
that Newtonian dynamics can be organized in the same self-consistent
fashion. Our main goal is to investigate the status of ergodic-nonergodic
transitions in dense liquids.  How does one approach the liquid-glass
transition?  In Ref. \cite{SDENE} the question of an ENE transition in
a system obeying SD by working at one-loop order was addressed.
With the simplest interaction vertices it was demonstrated
that a system of hard spheres does not under go an ENE transition
until a packing fraction of $\eta^{*} = 0.76$.  This is the reult of
the simplest calculation.
There are a variety of perturbations and extensions one can workout to test
the robustness of this result. A better description
for the static structure and/or improvements of
the vertex functions to include interaction effects can also be used in
the calculation. Going  to higher loop order is also another way
of improving the predictions of the theory.
We anticipate that all of these calculations will eventually
be carried out for SD.

There are differences between Newtonian and Smolukowsi
dynamics.  First, Newtonian dynamics are reversible and Smoluchowski
dynamics have a noise component.  Related is the fact that
ND has additional conservation laws (energy and momentum) compared
to SD. Finally there is the technically very important
fact that the phase-space in the ND case contains the particle
momenta while one has only the positions in the SD case.
In the SD case the kinetic equations lack the momentum
index labeling labeling ND kinetic equations. This is a large technical
advantage of working with the SD.
In the present work we take the point of view, which can be checked, that as
one approaches the glass transition the slow kinetics are associated
with structural rearrangements of the density, and couplings
to energy and momentum currents are less important.  This is
one of the assumptions of simplest version of the
mode-coupling theory\cite{MCT}.
A related working assumption is that the glassy kinetics of
SD and ND are very similar.  We know that they share the same static
correlation functions-static structure factors.
We will eventually be able to show this similarity.

An important formal point is that within the theory there exists
a {\bf core} problem involving $\rho$, the particle density, and
$B$, a response density. One must address
this core problem in both SD and ND before including other variables
like momentum variables in ND.  More specifically a discussion
of the rich hydrodynamical structure of this system is addressed
as one introduces additional variables.  For example the shear
viscosity is associated with the inclusion of transverse currents
in the development.

In the next section we show how kinetic theory can be reexpressed
in terms of path integrals.  The path integral formulation is similar
to that found in the SD case with the initial condition playing
a role similar to that for the noise in the SD case.
In the SD case\cite{FTSPD} the path integral approach involved the same
core fields:
$\rho$ and $B$.
The ND case can also be organized most simply in terms the same two
variables\cite{17}. With respect to their respective interaction structures,
the formulations are same for ND and SD dynamics.
We find that the noninteracting
cumulants for the set of variables $\rho$ and $B$
are qualitatively similar for the two cases.
Thus we  follow here the development
in Ref. \cite{FTSPD} and construct a self-consistent approach to the dynamics.

It has turned out in the SD case
that we can replace the bare potential with an effective potential
expressed in terms of the physical structure factor.  Thus the
theory can be applied to systems with hard-sphere interactions.
Here we develop a perturbation expansion for the two-point cumulant in terms of
the pseudo interaction potential.
The glue which ties together the terms in perturbation theory  are
the three-point vertices which are constructed from the
noninteracting three-point cumulants among the variables $\rho$ and $B$.
We show that all of these cumulants can be evaluated in the
time and wave number regime. We focus here on the two-point
cumulants and their determination to first order in the interaction
including the renormalization of the bare potential.

At the formal level we show that the collective contribution to the
self-energies for $\rho$ and $B$ have the same form in the ND case
as for the SD case to second order in the expansion. The main calculation carried out
in this paper is the determination of all the cumulants between
$\rho$ and $B$ for the noninteracting system. In the next paper
in this sequence we address the development of the theory at
second-order in the interaction.  The first important development
there is to show, more simply than in the SD case, that
one has a fluctuation dissipation symmetry relating the response
functions to the correlation functions.
In this case one can establish a set of nonperturbative
identities satisfied by the
three-point cumulants and irreducible vertices. Additional identities
are clearly available for higher order quantities.
In turn one shows that the set of matrix Dyson's equations, satisfied by the
two-point cumulants, reduce to a single
kinetic equation for the correlation function as assumed in mode-
coupling theory. It is from this kinetic equation that one can develop
the machinery associated with the ENE transition and the slow
dynamics one can observe even if one can not access a sharp
ENE transition.

In this paper we lay the ground work for the analogous calculations
for Newtonian dynamics. These calculations are in some ways harder
in the ND case. For example, the three-point vertex functions are
more complicated. However, in other ways the evaluations are
easier. The treatment of the FD symmetry is  simpler in ND case.
The expansion parameter in this problem is a pseudo-interaction
potential just as in Ref. \cite{SDENE} mentioned above.
Indeed we expect to find the same interaction
pseudo-potential in the two cases.

At second order we have an
expression for the static structure factor $S(q)$ in terms of the potential.  We
assume we can pick the best result for $S(q)$ and solve for the associated
pseudo-potential. Plots of $\tilde{V}(q)$ for various packing fractions
( {\em e.g.}, see eqn. (218) and Fig.2 in Ref. \cite{SDENE} )
shows  that the pseudo-potential is a rather smooth quantity even for rigid
hard spheres potentials.

A key difference between SD and ND is the noise in the SD case
driving the system toward equilibrium.  At low orders in the ND system
it needs to be told it is in equilibrium. In conventional kinetic theory
the initial state is chosen to be in equilibrium.  Once in equilibrium
the system typically stays there. Thus one option is to
fix an initial condition at $t=t_{0}$.
We show how this works in the zeroth and first-order cases for the
dynamic structure factor.  In the first-order case one has two effects on the theoretical
structure.  First one has (see Eq.(\ref{eq:27}) below) that the interaction
matrix has a piece which imposes an  initial condition and, second, the
time integrations are restricted to the time regime $t_{0}<t$.
We show that the fluctuation-dissipation theorem\cite{FDS} (FDT)
is crucial in seeing that these breaks in
time-translational invariance (TTI) cancel out and one obtains
correlation functions compatible with TTI.
In our treatment of FDT we find that the system is invariant over
a set of symmetry operations which depends on an undetermined
parameter.  We propose to fix the system at temperature $T$ by
choosing that all cumulants and vertices satisfy the FDS
associated with equilibrium.

\section{Newtonian Dynamics}

Consider a system of $N$ particles with mass $m$
 with configurations specified by the phase-space coordinates
$\Psi_{i}=(R_{i},P_{i})$
which satisfy the equations of motion:

\bea
\label{eq:2}
\dot{R}_{i} &=& \frac{P_{i}}{m} \\
\label{eq:3}
\dot{P}_{i} &=& f_{i}
\eea

\noindent where the particles experience force

\be
f_{i}=-\frac{\partial}{\partial R_{i}}U(R)
~~~,
\ee

\noindent with total potential energy

\be
U(R)=\frac{1}{2} \sum_{i\neq j}V(R_{i}-R_{j})
\ee

\noindent and we have suppressed vector labels to unclutter
the equations. If we form the vectors $\Psi_{i}=(R_{i},P_{i})$
then the equations of motion (\ref{eq:2}) and (\ref{eq:3})
can be put into the form

\be
\dot{\Psi}_{i}=K_{i}
\ee

\noindent where $K_{i}$ is a function of the $\Psi$.
We treat the phase-space coordinates
as our fields in a MSR structure\cite{MSR}
and the physical observables are
treated as conjugate to external fields which label the generators of
the physical observables. The $N$-particle partition function
is given by

\bea
\label{eq:a1}
Z_{N}[H,h,\hat{h}] &=& {\cal N}
\int \prod_{i=1}^{N}{\cal D}(\Psi_{i}){\cal D}(\hat{\Psi}_{i})
d\Psi_{i}^{(0)}) P_{0}(\Psi_{0} )
e^{-A_{\Psi}}
\nonumber \\
&\times& \exp(H\cdot \phi )\exp(h\cdot\Psi +i\hat{h}\cdot\hat{\Psi})
\label{eq:5}
\eea

\noindent where ${\cal N}$ is a normalization constant,
and we have an initial probability distribution $P_{0}(\Psi_{0} )$.
We assume the system is in equilibrium initially and the initial distribution is canonical:

\be
P_{0}=e^{-\beta {\cal H} (\Psi_{0})}/Z_{0}
\ee

\noindent where ${\cal H}$ is the hamiltonian, which is the sum of the kinetic
and potential energies.
The MSR action for the problem is given by

\be
A_{\Psi }=\int_{t_{0}}^{\infty}dt
\left[i
\hat{\Psi}_{i}(t)\cdot\left (\dot{\Psi}_{i}(t)+K_{i}(t)\right]\right) )
~~~.
\label{eq:9}
\ee

\noindent Finally we have the contribution in Eq.(\ref{eq:9})
 due to external fields that couple
to $\Phi$ the collective core variables of interest:

\be
\label{eq:Hf}
H\cdot \Phi =\sum_{\alpha}\int d1 H_{\alpha}(1)\Phi_{\alpha }(1)
~~~.
\ee

\noindent As explained in Ref. \cite{FTSPD}, the minimal  set for
$\Phi$ includes the
particle density and the response field $B$.
For convenience we have also included the fundamental source fields
$h_{i}$.
It is useful to show that this development can be mapped onto the traditional
representation.  Set $H=0$ in Eq.(\ref{eq:9}).  No information is lost as
long as we keep the full set of external fields $h$:

\be
\label{pfn1}
Z_{N}[h]={\cal N}\int {\cal D}(\Psi){\cal D}(\hat{\Psi})
d\Psi^{(0)}) P_{0}(\Psi_{0} )
e^{\left[i
\hat{\Psi}\cdot\left (\dot{\Psi}+K\right)\right]}
\exp(h\cdot\Psi )
\label{eq:2.9}
\ee

\noindent where in
each of the arguments of the exponentials on the RHS there is
 a summation over particle label,
an index labeling position and momentum, and an integral over time.
Now we do the functional integral over $\hat{\Psi}$ to obtain a functional
$\delta $-function:

\be
Z_{N}[h]={\cal N}\int {\cal D}(\Psi)
d\Psi^{(0)}) P_{0}(\Psi_{0} )
\delta \left( \dot{\Psi}+K\right)
\exp(h\cdot\Psi )~~~.
\label{eq:11}
\ee

\noindent The next step is to recognize that in a deterministic system the probability
of finding the system in configuration $\Psi (t)$ after starting at time
$t_{0}$ in configuration $\Psi_{0}$ is proportional to

\be
\delta\left( \Psi (t)-\Psi (t;t_{0})\right)
\ee

\noindent where $\Psi (t;t_{0})$ is the unique configuration at time $t$ evolving from
$\Psi (t_{0})$.  In the operator formulation

\be
\Psi (t;t_{0})=e^{i{\cal L}(t-t_{0})}\Psi (t_{0})
\ee

\noindent where ${\cal L}$ is the Liouville operator.
As discussed in some detail by Penco and Mauro\cite{penc},
one can use the following argument to connect our development here to the standard formulation.
If a function $f(\phi )$ has a zero at $\phi =\phi_{0}$ then

\be
\delta (\phi -\phi_{0})=\delta [f(\phi)]|f'(\phi_{0})|
\ee

\noindent In our case we choose $f$ to be $\dot{\Psi}+K$,
we have the identity

\be
\delta\left( \Psi (t)-\Psi (t;t_{0})\right)
=\delta\left(\dot{\Psi}(t)+K(t)\right){\cal N}
\label{eq:15}
\ee

\noindent and the factor ${\cal N}$ in Eqs.(\ref{eq:11}) and
(\ref{eq:15}) is the functional determinant

\be
{\cal N}=det\frac{\delta \left(\dot{\Psi}(t)+K(t)\right)}
{\delta \Psi (t')}
\ee

\noindent This quantity is essentially the Jacobian
discussed in Ref. \cite{FTSPD} which opens up
the discussion again to ghost fermions, supersymmetry,
unification\cite{gozz}. Importantly, for our purposes here,
${\cal N}$ is a
constant\cite{GRT} independent of $\Psi$. Using Eq.(\ref{eq:15})
in Eq.(\ref{eq:11})
leads to  the partition function as

\be
Z_{N}[h]=\int {\cal D}(\Psi)
d\Psi^{(0)} P_{0}(\Psi_{0} )
\prod_{t}\delta\left( \Psi (t)-\Psi (t;t_{0})\right)
exp(h\cdot\Psi )
~~~.
\ee

\noindent
We can immediately do the functional integral over $\Psi (t)$ to obtain

\be
\label{eq:19}
Z_{N}[h]=\int
d\Psi^{(0)} P_{0}(\Psi_{0} )
exp(\int_{t_{0}}^{\infty}dt\sum_{i=1}^{N}h_{i}(t)\cdot\Psi_{i} (t;t_{0}))
~~~.
\ee

\noindent Clearly by taking functional derivatives we can generate the average
of any set of phase-space observables:

\be
\langle A(t_{1})B(t_{2})\ldots D(t_{n})\rangle
=\int d\Psi^{(0)} P_{0}(\Psi_{0} ) A(\Psi (t_{1};t_{0}))
B(\Psi (t_{2};t_{0}))
\ldots
D(\Psi (t_{n};t_{0}))
~~~
\ee
and we see that our representation is equivalent to the standard theory.

\noindent The appropriate generating functional for the
problems discussed here, working in the grand canonical ensemble, is
given by

\be
W[H,h,\hat{h}]=ln Z_{T}[H,h,\hat{h}]
\label{eq:x}
\ee

\noindent where

\be
\label{eq:zT}
Z_{T}[H,h,\hat{h}]=\sum_{N=0}^{\infty}\frac{\rho_{0}^{N}}{N!}
Z_{N}[H,h,\hat{h}]
\ee

\noindent and $Z_{N}[H,h,\hat{h}]$ is given by Eq.(\ref{eq:19}).

For the method to be effective we have a minimum
of two  collective fields dictated by the structure
of the interactions.  One essential field
is the particle density

\be
\rho (1)=\sum_{i=1}^{N}\delta (x_{i}-R_{i}(t_{1}))
\ee

\noindent
and  it is crucial to include the response field

\be
B(1) =\sum_{i=1}^{N}\left[(\hat{P}_{i}(t_{1}))i\nabla_{1}
\right]
\delta (x_{1}-R_{i}(t_{1}))
~~~.
\ee

\noindent Notice that $B$ depends on the MSR hatted field $\hat{P}_{i}(t)$.
Unlike the SD case the {\it Jacobian} does not contribute to the definition
of $B(1)$ here.

We can then write the canonical partition function given by
Eq.(\ref{eq:a1}) in the form
\be
Z_{N}[H,h,\hat{h}]
=\int \prod_{i=1}^{N}{\cal D}(\Psi_{i}){\cal D}(\hat{\Psi}_{i})
{\cal D}(\Psi_{i}^{(0)})
e^{-A_{0}-A_{I}+H\cdot\phi+h\cdot \psi +\hat{h}\cdot\hat{\psi}}
\label{eq:24}
\ee

\noindent where $A_{0}$ is the quadratic part of the action
including the quadratic contribution to the
initial probability distribution

\be
A_{0 }=\int_{t_{0}}^{\infty}dt_{1}
\left[\sum_{i=1}^{N}
i\hat{\Psi}_{i}\cdot\left (\dot{\Psi}_{i}+K^{(0)}_{i}\right)\right]
+\beta K_{0}~~.
\label{eq:2.20}
\ee

\noindent $K_{0}$ is the initial kinetic energy.
Notice that we have constructed things such that the phase-space
variables are constrained to their initial values at
$t=t_{0}$.  We then average over these values.  Here we are
explicitly treating the case where the system is in equilibrium
at $t=t_{0}$, but more general situations are clearly compatible
with the development.
The interaction part of the action is given in the compact form

\be
A_{I}=\frac{1}{2}\sum_{\alpha,\nu}\int d1d2
\Phi _{\alpha}(1)\sigma _{\alpha\nu} (12)\Phi _{\nu}(2)
\label{eq:26}
\ee

\noindent where the Greek labels range over  $\rho$ and $B$ and the
interaction matrix is defined just as in SD, by

\be
\sigma_{\alpha \beta} (12)=V(12)
\left[ \delta_{\alpha\rho}\delta_{\beta\rho}
\beta \delta (t_{1}-t_{0})
+\delta_{\alpha{B}}\delta_{\beta\rho }
+\delta_{\alpha\rho}\delta_{\beta{B}}\right]
\label{eq:27}
\ee

\noindent where the first contribution is from the potential
energy contribution to the initial condition
and

\be
V(12)=V(x_{1}-x_{2})\delta (t_{1}-t_{2})
~~~.
\ee

\noindent Notice that the
 response field $B$ is chosen such that the interaction
part of the action has the form given by Eq.(\ref{eq:26}).
The canonical partition function (\ref{eq:24})  can be written in the
convenient form

\be
Z_{N}=\tilde{Tr} e^{-A_{I}+H\cdot\Phi}
\label{eq:23}
\ee

\noindent where we have introduced the average

\be
\tilde{Tr} =\int \prod_{i=1}^{N}{\cal D}(\Psi_{i}){\cal D}(\hat{\Psi}_{i})
d\Psi_{i}^{(0)} e^{-A_{0}}
~~~.
\ee

We have thus shown that ND can be written in a path-integral form as given in
Eqns.(\ref{eq:23a})-(\ref{eq:23c}). These expressions look like the formulation for SD with an
important exception.  The action $A_{\psi}$, given by Eq.(\ref{eq:9}),
does not have a contribution from noise.  A noise component, as in
Fokker-Planck dynamics, adds a term to the action

\be
A_{noise}=\int_{t_{0}}^{\infty}dt_{1}
\int_{t_{0}}^{\infty}dt_{2}\sum_{i=1}^{N}\hat{\psi}_{i}(t_{1})
D(t_{1},t_{2})\hat{\psi}_{i}(t_{2})
\ee

\noindent which contributes to the noninteracting part of the action
Eq.(2.20).  The noise has the property of continuously telling
the system to equilibrate at temperature $T$.  How in the case
of Newtonian reversible dynamics does the system know it is in equilibrium?
One mechanism is to satisfy an initial condition.
In the conventional formulation an equilibrium
correlation function is given by
\be
C_{AB}(t)=Tr P_{0}Be^{i{\cal L}t}A
\ee
where ${\cal L}$ is the Liouville operator and $P_{0}$ the
equilibrium probability distribution

\noindent
and at $t=0$ one has explicitly an equilibrium probability distribution.
The use of initial conditions is one important way of treating
nonequilibrium kinetics as discussed in Ref. \cite{FTSPD}.
The use of initial conditions in the case where the system is in equilibrium
for all times, including the $t=t_{0}$, is inconvenient since formally
it looks like one has broken time-translational invariance (TTI).
As shown below for an ideal gas and to first order in the interaction,
one can {\it tell} the gas its in equilibrium with an initial
condition in equilibrium and maintaining TTI.
However there is a cleaner way of maintaining equilibrium and TTI.
This is to require that the fluctuation-dissipation symmetry hold at all times.

\subsection{Fluctuation-Dissipation relations}

The correlation functions, via its definition as the average of the
product of commuting (classical) fields satisfies the time reversal
symmetry

\be G_{i  j}(t-t')=G_{ ji }(t'-t) \nonumber \ee \noindent The
physical fields are real so $G_{ij}^{*}(t-t')=\langle
\psi_i(t)\psi_j(t')\rangle^{*}= G_{i j}(t-t')$. In a stationary
state we have time translational invariance. The full MSR action as
out lined above  is obtained in the following form for the Newtonian
Dynamics

\bea \label{action-e1} {\cal A} &=& \sum_i \int \Bigg [
i\hat{R}_i(t) \Big \{ \dot{R}_i(t)-\frac{P_i(t)}{m} \Big \} +
i\hat{P}_i(t)\Big \{
\dot{P}_i(t)- F_i(t) \Big \} \Bigg ] dt \nonumber \\
\eea

\noindent Under complex conjugation the MSR action transforms as

\be {\cal A}^{*}(\psi ,\hat{\psi})= {\cal A}(\psi ,-\hat{\psi})~~.
\label{eq:54} \ee

\noindent We introduce
the following transformation:

\bea
\tau R_i(t) &=& R_i(-t) \nonumber \\
\tau\hat{R}_i(t) &=& -\hat{R}_i(-t) + i{\beta}F_{i}(-t)\nonumber \\
\tau P_i(t) &=& P_i(-t) \nonumber \\
\tau\hat{P}_i(t) &=& -\hat{P}_i(-t) - \frac{i\beta}{m}P_i(-t)
\label{ABL-trans} \eea

\noindent where $F_i(t)=({\partial}U/{\partial}R_i(t))$ is the force
on the particle $i$. We now consider how the action ${\cal A}$
changes under this transformation

\bea
{\cal A}^{'} &=& \tau {\cal A} \nonumber \\
&=& \sum_i \int dt \Bigg [ -i\hat{R}_i(-t) \left \{
\frac{{\partial}R_i(-t)}{\partial(-t)}-\frac{P_i(-t)}{m} \right \}
-i\hat{P}_i(-t)\left \{
\frac{{\partial}P_i(-t)}{\partial(-t)}-F_i(-t) \right \}
\Bigg ] \nonumber \\
&+&  {\beta}\int_{-\infty}^{+\infty}
\frac{\partial{H(-t)}}{\partial(-t)}d(t) \nonumber \eea

\noindent Looking at individual terms, and letting $t\rightarrow -t$
in the integrals, we obtain
$A'=\tau{A}=A-{\beta}H(-\infty)+{\beta}H(+\infty)$. We take the two
limits of time integration $t_{2}\rightarrow\infty$ and
$t_{1}\rightarrow -\infty$ here. Treating the last part as a
constant (=0 in a conserved case) we conclude that the MSR action
remains invariant under this transformation.

\subsection{FDT involving the $\rho$ and $B$ fields}

Let us consider the transformation rule for the field $B(x,t)$ under
time reversal $\tau$

\begin{equation}
\label{Bfdt1} B(x,t)=-\sum_{i=1}^N \hat{\vec P}_i \cdot
\frac{\partial}{\partial{{\vec R}_i}}\delta(x-R_i(t))
\end{equation}

\noindent In the present formulation we are working so far with the
density variable $\rho(x,t)$ as the only collective variable and
$B(x,t)$ is the hatted counterpart. Under $\tau$ the field $B(x,t)$
changes as

\begin{eqnarray}
\label{Bfdt3} \tau B(x,t) &=& -B(x,-t)-\frac{i}{m}{\beta}\Big [
\frac{\partial\rho(x,t)}{\partial(t)} \Big ] \nonumber
\end{eqnarray}

\noindent Using this relation we obtain for any function $f[\rho]$
the following FDT relation linking to the $B(x,t)$

\begin{equation}
\label{Bfdt4} G_{fB}(t-t')=\frac{i}{m}\theta
(t-t'){\beta}\frac{\partial}{\partial{t}}{G_{f\rho}}(t-t')
\end{equation}

\section{Self-Consistent Development}

The self consistent theory for the liquid state dynamics is developed in
terms of correlation functions of collective variables
$\Phi_\alpha$'s introduced
in the previous section. We use the notation for the
variable $\Phi_\alpha$ as

\be
\lbl{eq:Phidef}
\Phi_\alpha = \sum_i \phi^{(i)}_\alpha
\ee

\noindent with the index $\alpha$ denoting the space of collective
variables and $i$ is the particle label.
We work here with the density variable $\rho$ and
conjugate variable $B(x,t)$ which are respectively denoted as:

\be
\phi^{(i)}_{\rho}(1)=\delta (x_{1}-R_{i}(t_{1}))~~
\ee

and

\be
\phi^{(i)}_{B}(1)=i\hat{P}_{i}(t_{1})\nabla_{x_{1}}
\delta (x_{1}-R_{i}(t_{1}))
~~~.
\ee

\noindent Working in the grand canonical
ensemble, the grand  partition function $Z_T$ for the interacting problem
is given by Eq. (\ref{eq:zT}).
The cumulants of the fields $\Phi_{i}$  are generated by
taking functional derivatives of the generating functional

\be
W[H]=ln~ Z_{T}
~~~.
\ee

\noindent with respect the fields $H$ introduced in eqn. (\ref{eq:Hf}) above.
The one-point average in a field is given by

\be
G_{i}=\frac{\delta}{\delta H_{i}}W
\nonumber
\ee

\noindent In the above equation we have used a compact notation where
the index $i$ labels space, time and fields $\rho$ or $B$.
We maintain the notation from  here on.
In Ref. \cite{FTSPD} we derived the fundamental identity

\be
G_{i}
=Tr \phi_{i} e^{H\cdot\phi +\Delta W}
\label{eq:32}
\ee

\noindent where

\be
\Delta W =W[H+F]-W[H]
\ee

\noindent with $H$ denoting the external field and $F_i$ being given by

\be \label{fi-core}
 F_{i}=\sum_{j}\sigma_{ij}\phi_{j} ~~~.
\ee

\noindent with the interaction kernel $\sigma_{ij}$ obtained in Eq.
(\ref{eq:27}). The important
 result Eq. (\ref{eq:32}) was established in FTSPD.
  It is more useful to derive
here this result in a completely different fashion. This is presented in
Appendix \ref{appA} in which we derive Eq.(\ref{eq:32}) without using functional
techniques.

\section{Formal Development of Perturbation Theory}

The dependence of the theory  on the pair potential is
controlled by the quantity

\be
\label{eq:delW}
\Delta W=W[H+F]-W[H]
\ee

\noindent  in eqn.(\ref{eq:32})
where $F$ is proportional to the interaction potential.
We can expose the dependence on the potential by
constructing the functional Taylor-series expansion

\be
\label{delw-exp}
\Delta W =\sum_{i}F_{i}\frac{\delta}{\delta H_{i}}W[H]
+\sum_{ij}\frac{1}{2}F_{i}F_{j}\frac{\delta^{2}}{\delta H_{i}\delta H_{j}}W[H]+\cdots
\ee

\noindent and we can introduce the full cumulants:

\be
G_{ij\ldots k}=\frac{\delta}{\delta H_{i}}\frac{\delta}{\delta H_{j}}
\ldots \frac{\delta}{\delta H_{k}}
W[H]
\ee

\noindent to obtain

\be
\Delta W =\sum_{i}F_{i}G_{i}
+\sum_{ij}\frac{1}{2}F_{i}F_{j}G_{ij}
+\sum_{ijk}\frac{1}{3!}F_{i}F_{j}F_{k}G_{ijk}
+\ldots
~~~.
\ee

\noindent Clearly in this form we can take $\Delta W$ to be a functional of $G_{i}$.
One can then use functional differentiation to express higher
order cumulants in terms of  $G_{i}$ and
$G_{ij}$ .  One has for example the
manipulation expressing the three-point cumulant in terms
of $G_{ij}$

\be
G_{ijk}=\frac{\delta}{\delta H_{k}}G_{ij}
=\sum_{mnp} -G_{im}G_{jn}G_{kp}
\Gamma_{mnp}
\ee

\noindent and the irreducible three-point vertex. $\Gamma_{ijk}$ in turn
is given as a functional
derivative of the two-point irreducible vertex,

\be
\lbl{eq:t-vert}
\Gamma_{ijk}=\frac{\delta}{G_{k}}\Gamma_{ij}
\ee

\noindent with respect to $G_{k}$.
$\Gamma_{ij}$ is precisely the matrix inverse of
the two-point cumulant $G_{kj}$:

\be
\lbl{eq:sedef}
\sum_{k}\Gamma_{ik}
G_{kj}=\delta_{ij}
~~~
\ee
where we refer to this as Dyson's equation.

\noindent As in the SD case \cite{SDENE} we can establish
a dynamic generalization of the static Ornstein-Zernike
relation\cite{OZR}.  Starting with the functional equation
for the two-point cumulant, we use the chain-rule for functional
differentiation to obtain:

\bea
G_{ij} &=& \frac{\delta}{\delta H_{j}}G_{i} \nonumber \\
&=& Tr \phi_{i}\phi_{j}e^{H\cdot\phi +\Delta W}
+\sum_{k}Tr \left( \phi_{i}\frac{\delta}{\delta G_{k}}e^{H\cdot\phi+\Delta W} \right)
\frac{\delta}{\delta H_{j}}G_{k} \nonumber \\
&=& {\cal G}_{ij}+\sum_{k}c_{ik}G_{kj}
\label{eq:461}
\eea

\noindent where

\be
{\cal G}_{ij}=Tr \phi_{i}\phi_{j}e^{H\cdot\phi +\Delta W}
\ee

\noindent is roughly speaking a one-body object, and $c_{ij}$ is defined as

\be
c_{ij}=Tr \phi_{i}e^{H\cdot\phi +\Delta W}
\frac{\delta}{\delta G_{k}}\Delta W
~~~.
\label{eq:63}
\ee

\noindent Since $\Delta W$ can be treated as a functional of $G_{i}$ we see
at this stage that we have available a self-consistent theory.
To formulate this  we define the matrix-inverse of the one body quantity
${\cal G}_{ij}$ as $\gamma_{ij}$,

\be
\lbl{eq:gamdef}
\sum_{k}\gamma_{ik}{\cal G}_{kj}=\delta_{ij}
~~~.
\ee

\noindent
Now multiplying the correlation matrix $G$ defined in
eqn. (\ref{eq:461}) with the matrix $\gamma$, and using
definitions (\ref{eq:sedef}) and
(\ref{eq:gamdef}), we obtain the two-point vertex function
(without any approximation):

\be
\Gamma_{ij}=\gamma_{ij}+K_{ij}
\label{eq:50}
\ee

\noindent where the collective part or the dynamic self-energy is given by

\be
K_{ij}=-\sum_{k}\gamma_{ik}c_{kj}
\label{eq:501}
~~~.
\ee

\subsection{Collective Self-Energy at First and Second Order}

At the formal level we can work out the collective part of
the self-energy in perturbation theory.  Using Eq.(\ref{eq:63}) we have
at first order

\be
c_{ij}^{(1)}=Tr \phi_{i}e^{H\cdot\phi +\Delta W}
\frac{\delta}{\delta G_{j}}\Delta W^{(1)}
~~~.
\label{eq:48}
\ee

\noindent where

\be
\lbl{dw1}
\Delta W^{(1)}=\sum_{i}F_{i}G_{i}
~~~.
\ee

\noindent  Clearly

\be
\frac{\delta}{\delta G_{j}}\Delta W^{(1)}=F_{j}
\ee

\noindent and

\bea
c_{ij}^{(1)} &=& Tr \phi_{i}e^{H\cdot\phi +\Delta W} F_{j} \nonumber \\
&=& Tr \phi_{i}e^{H\cdot\phi +\Delta W} \sum_{k}\phi_{k}\sigma_{kj}
\nonumber \\
&=& {\cal G}_{ik}\sigma_{kj}
~~~.
\lbl{eq:c1}
\eea

\noindent The collective contribution to the self-energy at first-order is

\be
\lbl{eq:mem1}
K_{ij}^{(1)}=-\gamma_{i\ell}{\cal G}_{\ell k}\sigma_{kj}
=-\sigma_{ij}
~~~.
\ee

\noindent Next, we consider the second-order contribution

\be
c_{ij}^{(2)}=Tr \phi_{i}e^{H\cdot\phi +\Delta W}
\frac{\delta}{\delta G_{j}}\Delta W^{(2)}
\ee

\noindent where

\be
\Delta W^{(2)}=\frac{1}{2}\sum_{ij}F_{i}F_{j}G_{ij}
~~~.
\ee

\noindent It is shown in Ref. \cite{SDENE} that the second order contribution
to the dynamic self energy is obtained in the symmetric form in terms of the
screened matrix correlation function $\bar{G}$ as

\be
\lbl{eq:bubble}
K_{ij}^{(2)}=-\frac{1}{2}\gamma_{iuv}^{(0)}\bar{G}_{ur}\bar{G}_{vq}
\gamma_{jrq}^{(0)}
\ee

\noindent where $\gamma^{(0)}_{ijk}$ denotes the three point vertex function
$\Gamma_{ijk}$ defined in eqn. (\ref{eq:t-vert}) at the lowest order {\em i.e.},
for a noninteracting system.
This is referred to as the collective one-loop contribution.

The $\bar{G}$ matrix elements are obtained as

\be
\bar{G}_{ij}(1)=\frac{1}{2}\sum_{kl} \left [ G^{(0)}_{ik}
\sigma_{kl}G_{lj}+{G}_{ik}\sigma_{kl}G_{lj}^{(0)}\right ]
\ee

\noindent and plays the role of  an effective propagator.
The above result obtained for the
SD case in Ref. \cite{SDENE} also holds for the ND case. The relations
(\ref{eq:50}) and (\ref{eq:bubble}) form a closed set of
self-consistent equations for the correlation matrix $G_{ij}$.
This  gives rise to a feed back mechanism which becomes
strong enough at high density to drive the system to a
possible ENE transition. Evaluating the strength of the feed back mechanism
even at the lowest order will involve evaluating the
three- point vertex functions
$\gamma^{(0)}_{ijk}$ for the noninteracting system. Later we describe the
basic calculation for obtaining the correlations in the ideal gas.
Analysis of the feed back process with evaluation of the vertex functions
and the possibility of the ENE transition will be taken up in a companion paper.

\subsection{First-Order Theory}

To obtain the correlation functions at the first order we need as
an input the correlation functions for the noninteracting system.
We present computation of the two point correlation functions
$G_{\alpha\beta}^{(0)}$ at the zeroth order in the next section. We
list the zeroth- order results here as a starting point of the first- order
calculation :

\bea
\lbl{tpt0-rr}
G_{\rho\rho}^{(0)}(k,t_1,t_2)&=& \rho_0 e^{-\frac{k^2}{2\beta{m}}{(t_1-t_2)}^2} \\
\lbl{tpt0-rb}
G_{\rho{B}}^{(0)}(k,t_1,t_2)&=& -\frac{k^2}{m}(t_1-t_2)
e^{-\frac{k^2}{2\beta{m}}{(t_1-t_2)}^2}\theta(t_1-t_2)\rho_{0}\\
\lbl{tpt0-br}
G_{B\rho}^{(0)}(k,t_1,t_2)&=& -\frac{k^2}{m}(t_2-t_1)
e^{-\frac{k^2}{2\beta{m}}{(t_1-t_2)}^2}\theta(t_2-t_1)\rho_{0}\\
\lbl{tpt0-bb}
G_{BB}^{(0)}(k,t_1,t_2 ) &=& 0
\eea

\noindent where we have the same structure as in the SD case.
$G_{\rho\rho}$ is real, $G_{\rho B}$ is retarded, $G_{B\rho}$ is
advanced, and $G_{BB}$ is zero. For the time translational
invariance  we obtain the frequency transformed quantities as

\bea
\lbl{tptf0-rr}
G_{\rho\rho}^{(0)}(k,\omega)&=& \sqrt{2\pi}\frac{\rho_0}{kv_0}
e^{-\frac{\omega^2}{2k^2v_0^2}} \\
\lbl{tptf0-rb}
G_{\rho{B}}^{(0)}(k,\omega)&=& -\rho_0\beta {\cal S}
\Big ( \frac{\omega}{\sqrt{2}kv_0} \Big )
\eea

\noindent
where $mv_{0}^{2}=k_BT$ and
where the integral ${\cal S}(x)$ is defined as

\be
\lbl{def:i1}
{\cal S}(x) = 1-2x^2e^{-x^2}\int_0^x du e^{u^2}+i\sqrt{\pi}xe^{-x^2}
\ee

Let us look at the first-order theory for the one point function
or the equation of state. Starting from the basic relation (\ref{eq:32})
we obtain for the one point function

\bea
G_i &=& G^{(0)}_i + Tr \phi_i \Delta W^{(1)} \nonumber \\
    &=& G^{(0)}_i+G^{(0)}_{ik}\sigma_{kl}G_l
\label{eq:321}
\eea

\noindent For $i=\rho$ using the corresponding $\sigma_{B\rho}$ element
and using the result $G_\rho(1)=2\pi\delta(\omega_1)G_\rho(k_1)$
we obtain

\be
\lbl{rho1}
G_\rho(k)=\frac{G^{(0)}_\rho(k)}{1 - V(k)G^{(0)}_{\rho{B}}(k,0)}
\ee

\noindent Since $G_\rho^{(0)}(0) \equiv \rho_0$ we obtain the equation of state
to first order in $V$   as

\be
\lbl{rhobar}
\bar{\rho}=\frac{\rho_0}{1 + \beta{\rho_0}V}
\ee

\noindent where we have used the result from eqn. (\ref{tptf0-rb})
that $G_{\rho{B}}^{(0)}(0,0)=-\beta\rho_0$ to zeroth order in the interaction.
Notice that we self-consistently determine the static properties
as we solve the dynamic problem.

It turns out to be more useful to write eqn. (\ref{rhobar}) in the form:

\be
\rho_{0}=\frac{\bar{\rho}}{1-\beta\bar{\rho}V(0)}
\nonumber
\ee
\be
=\bar{\rho}exp(\tilde{V}(0))
~~~.
\ee

\noindent We now relate this to the conventional equation of state.
First, since we are working in the grand canonical ensemble,
$\rho_{0}$ is the fugacity
\be
\rho_{0}=\frac{e^{\beta \mu}}{\ell^{d}}
\ee

\noindent where $\ell$ is a microscopic length independent of density and $\mu$
is the chemical potential.  There is the Gibbs-Duhem relation

\be
\frac{\partial p}{\partial \bar{\rho}}=\bar{\rho}
\frac{\partial \mu}{\partial \bar{\rho}}
\ee

\noindent or

\be
\frac{\partial \beta p}{\partial \bar{\rho}}=\bar{\rho}
\frac{\partial \beta \mu}{\partial \bar{\rho}}
\ee
\be
=\frac{\bar{\rho}}{\rho_{0}}
\frac{\partial \rho_{0}}{\partial \bar{\rho}}
~~~.
\ee
\noindent Working at first order in the potential it is easy to show
\be
\frac{\partial \rho_{0}}{\partial \bar{\rho}}
=e^{\tilde{V}[0]}(1+\tilde{V}[0])
\ee
\noindent Then
\be
\frac{\partial \beta p}{\partial \bar{\rho}}=
(1+\tilde{V}[0])
\ee
\be
\beta p = \bar{\rho}+\frac{1}{2}\beta\bar{\rho}^{2}V[0]
\ee
\noindent The first term is the ideal gas law. The complete expression
with $V$ an effective potential is sensible over a broad range of
densities.

Let us look at the first-order theory for the two-point
correlation functions. To obtain
the correlation functions at this order we begin with
the eqn. (\ref{eq:461}) and use the corresponding order
expression for the self energy $c^{(1)}_{ij}$ and $\Delta W^{(1)}$
obtained respectively in eqns. (\ref{eq:c1}) and (\ref{dw1})
to write the matrix kinetic equation\cite{FTSPD} :

\be
G_{ij} = G^{(0)}_{ij}+\sum_{l,k}G_{ijl}^{(0)}\sigma_{lk}G_k
+\sum_{l,k} G^{(0)}_{il} \sigma_{lk}G_{kj}
\label{eq:462}
\ee

\noindent The role of the second  term on the RHS of the above equation gives rise to a factor
\cite{FTSPD} of $\bar{\rho}/\rho_0$ with the first term.  In time space we write,

\be
\label{eq:80}
G_{ij}(k,t_1,t_2) = \frac{\bar{\rho}}{\rho_0} G^{(0)}_{ij}(k,t_1,t_2)+
\int_{t_{0}}^{\infty}d\bar{t}\sum_{k,l}
G^{(0)}_{ik}(k,t_1,\bar{t}) \sigma_{kl}(k) G_{lj}(k,\bar{t},t_2)
~~~.
\ee

\noindent The average density $\bar{\rho}$ is defined in the equation of state
given by  eqn. (\ref{rhobar}) in the time translational invariant  form.
Now from eqn.(\ref{eq:80}) we obtain setting $i=j=B$ and $t_2=t_0$
(the initial time)

\bea
G_{BB}(k,t_1,t_2) &=& \frac{\bar{\rho}}{\rho_0} G^{(0)}_{BB}(k,t_1,t_2) +
G^{(0)}_{B\rho}(k,t_1,t_0) (-\beta{V(k)})G_{\rho{B}}(k,t_0,t_2) \nonumber \\
~~~.
\eea

\noindent Since $G^{(0)}_{BB}(k,t_1,t_2)$, $G_{B\rho}^{{(0)}}(k,t_1,t_0)$
and $G^{(0)}_{\rho{B}}(k,t_0,t_2)$ are zero,
we find $G_{BB}(k,t_1,t_2)=0$.
Consider the response functions
$G_{\rho{B}}(k,t_1,t_2)$ and $G_{B\rho}(k,t_1,t_2)$ by taking
the $\rho -B$ and $B-\rho$ matrix elements of Eq.(\ref{eq:80})
results in the equations

\bea
G_{\rho{B}}(k,t_1,t_2) &=& \frac{\bar{\rho}}{\rho_0} G^{(0)}_{\rho{B}}(k,t_1,t_2)+
\int_{t_{0}}^{\infty}d\bar{t}G^{(0)}_{\rho{B}}(k,t_1,\bar{t})
V(k)G_{\rho{B}}(k,\bar{t},t_2) \nonumber \\
G_{B\rho}(k,t_1,t_2) &=& \frac{\bar{\rho}}{\rho_0}
G^{(0)}_{B\rho}(k,t_1,t_2) + \int_{t_{0}}^{\infty}d\bar{t}
G^{(0)}_{B\rho}(k,t_1,\bar{t})V(k)G_{B\rho}(k,\bar{t},t_2) \nonumber \\
\label{eq:81}
\eea

\noindent It is easy to see that
$G_{\rho{B}}(k,t_0,t_2)$ and $G_{B\rho}(k,t_1,t_0)$ are both equal to zero.
Since $G_{\rho B}(k,\bar{t},t_{2})$ is retarded and
$G_{B\rho }^{(0)}(k,t_{1},\bar{t})$ is advanced,
eqn.(\ref{eq:81}) are solved by Fourier transformation:

\bea
\label{eq:84a}
G_{\rho{B}}^{(1)}(k,\omega ) &=&
\frac{({\bar{\rho}}/{\rho_0})G^{(0)}_{\rho{B}}(k,\omega )}
{1-V(k)G^{(0)}_{\rho{B}}(k,\omega )} \nonumber \\
G_{B\rho}^{(1)}(k,\omega ) &=& \frac{({\bar{\rho}}/{\rho_0})G^{(0)}_{B\rho}(k,\omega )}
{1-V(k)G^{(0)}_{B\rho}(k,\omega )} \nonumber \\
\eea

\noindent Using the results from eqn. (\ref{tptf0-rb})
that $G_{\rho{B}}^{(0)}(0,k)=-\beta\rho_0$, to zeroth order in the interaction,
we obtain the zero-frequency $G_{\rho{B}}$ at the first order as,

\bea
\label{eq:84aa}
G_{\rho{B}}^{(1)}(k,0) &=&
-\frac{\beta\bar{\rho}}{1+\beta\rho_0V(k)}\nonumber \\
G_{B\rho}^{(1)}(k,0) &=&  -\frac{\beta\bar{\rho}}{1+\beta\rho_0V(k)}
\nonumber \\
\eea

\noindent
Now we consider the $\rho -\rho$ matrix element of Eq.(\ref{eq:80})

\bea
G_{\rho\rho}(k,t_1,t_2) &=& \frac{\bar{\rho}}{\rho_0} G^{(0)}_{\rho\rho}(k,t_1,t_2) +
G^{(0)}_{\rho\rho}(k,t_1,t_0) (-\beta{V(k)})G_{\rho\rho}(k,t_0,t_2)
\nonumber \\ &+& \int_{t_{0}}^{\infty}d\bar{t} G^{(0)}_{\rho\rho}(k,t_1,\bar{t})
V(k)G_{B\rho}(k,\bar{t},t_2) \nonumber \\
&+& \int_{t_{0}}^{\infty}d\bar{t}
G^{(0)}_{\rho{B}}(k,t_1,\bar{t})V(k)G_{\rho\rho}(k,\bar{t},t_2)
\eea

\noindent We rewrite this equation in the following form

\bea
G_{\rho\rho}(k,t_1,t_2) &=& \frac{\bar{\rho}}{\rho_0} G^{(0)}_{\rho\rho}(k,t_1,t_2)
+ \int_{-\infty}^{\infty}d\bar{t} G^{(0)}_{\rho\rho}(k,t_1,\bar{t})
V(k)G_{B\rho}(k,\bar{t},t_2) \nonumber \\
&+& \int_{-\infty}^{\infty}d\bar{t}
G^{(0)}_{\rho B}(k,t_1,\bar{t})V(k)G_{\rho\rho}(k,\bar{t},t_2)
+ K(k,t_{1},t_{2}) \nonumber \\
\label{eq:84}
\eea

\noindent where all of the terms dependent on $t_{0}$ are included in
\bea
K(k,t_{1},t_{2}) &=&
G^{(0)}_{\rho\rho}(k,t_1,t_0) (-\beta{V(k)})G_{\rho\rho}(k,t_0,t_2)
\nonumber \\
&-& \int_{-\infty}^{t_{0}}d\bar{t} G^{(0)}_{\rho\rho}(k,t_1,\bar{t})
V(k)G_{B\rho}(k,\bar{t},t_2) \nonumber \\
&-& \int_{-\infty}^{t_{0}}d\bar{t}
G^{(0)}_{\rho{B}}(k,t_1,\bar{t})V(k)G_{\rho\rho}(k,\bar{t},t_2)
~~~.
\label{eq:85}
\eea

All of the elements contributing to the breaking TTI are collected
into $K$.  By direct calculation one can show that $K=0$ and
TTI holds.  This analysis is rather involved.  Instead we
consider the case where the FDT holds

\be
G_{\rho B}(t-s)=\theta (t-s)\beta
\frac{\partial}{\partial t}G_{\rho\rho}(t-s)
~~~.
\ee

\noindent Now in $K$ eliminate the response functions in terms of the density-density
correlation function.  Notice that the two terms involving
an integration can be combined into a single term where inside the
integral one has an exact derivative.  Doing the integration
the contribution from $\bar{t}=t_{0}$ cancels the term from the
initial condition and one has $K=0$. Eqn.(\ref{eq:84}) with $K=0$ can be solved by Fourier transformation
and one has

\be
G_{\rho\rho}=\frac{(\bar{\rho}/\rho_0)G_{\rho\rho}^{(0)}}{1-V(k)G_{\rho B}^{(0)}}
\left[1+\frac{V(k)G_{B\rho}^{(0)}}
{1-V(k)G_{B\rho }^{(0)}}\right]
\ee

\noindent and we have used the previously determined result for the response
functions. After a little rearrangement we have the real result

\be
G_{\rho\rho}=\frac{(\bar{\rho}/\rho_0)G_{\rho\rho}^{(0)}}
{[1-V(k)G_{\rho B}^{(0)}]
[1-V(k)G_{B\rho }^{(0)}]}
\ee

\noindent where

\be
G_{\rho B}^{(0)}
=\left[G_{B\rho }^{(0)}\right]^{*}
~~~.
\ee

\noindent This is the solution to our first order problem.

There are other ways of looking at the information contained in the solution.
If the FDT holds we have in Fourier space

\be
\lbl{fdtreln}
\frac{2}{\beta\omega}Im G_{\rho B}=G_{\rho\rho}
\ee

\noindent which also holds for the noninteracting limit

\be
\lbl{fdtreln0}
\frac{2}{\beta\omega}Im G^{(0)}_{\rho B}=G^{(0)}_{\rho\rho}
\ee

\noindent These lead to an integral form for the FDT.  We start with the identity
for the Laplace transform of $G_{\rho{B}}$

\bea
G_{\rho B}(z) &=& \int_{-\infty}^{\infty}\frac{d\omega}{\pi}
\frac{Im G_{\rho B}}{z-\omega}
\nonumber \\
&=& \int_{-\infty}^{\infty}\frac{d\omega}{2\pi}
\beta\omega\frac{G_{\rho \rho}}{z-\omega}
\nonumber \\
&=& \beta \int_{-\infty}^{\infty}\frac{d\omega}{2\pi}
G_{\rho \rho}(\omega )
\frac{\omega -z +z}{z-\omega}
\nonumber \\
&=& \beta \int_{-\infty}^{\infty}\frac{d\omega}{2\pi}
G_{\rho \rho}(\omega )[-1+\frac{z}{z-\omega}]
\nonumber \\
&=& -\beta \bar{\rho}S(k)
+z\beta G_{\rho \rho} (k,z)
\label{eq:75}
\eea

\noindent
where the Laplace transform of the density-density correlation
function is defined as

\be
G_{\rho\rho}(k,z)
=\int_{-\infty}^{\infty}\frac{d\omega}{2\pi}
\frac{G_{\rho \rho}(k,\omega)}{z-\omega}
~~~.
\ee

\noindent  We have used above the result

\bea
G_{\rho B}(z=0) &=&  -\beta \int_{-\infty}^{\infty}\frac{d\omega}{2\pi}
G_{\rho \rho}(\omega ) \nonumber \\
&=&  -\beta \bar{\rho} S(k)~~,
\eea

\noindent where $S(k)$ is the static structure factor.
Using eqn. (\ref{eq:84aa})  we obtain our first order
approximation for the structure factor

\be
S(k)=\frac{1}{1+\rho_{0}\beta V(k)}
\ee

\noindent which is equivalent to taking the
two point Ornstein-Zernike direct correlation function as

\be
c(k)=-\beta V(k)~~.
\ee

\noindent
We then equate the two expressions for $G_{\rho B}(k,z)$ obtained in
eqns. (\ref{eq:84a}) and  (\ref{eq:75})

\bea G_{\rho B}(z) &=& -\beta \bar{\rho}\rho S(k)
+z\beta G_{\rho \rho} (k,z) \nonumber \\
&=& \frac{(\bar{\rho}/\rho_0)G^{(0)}_{\rho{B}}(k,z)}
{1-V(k)G^{(0)}_{\rho{B}}(k,z )}
~~~.
\eea

\noindent Solving for $G_{\rho\rho}$ we find
\be
\lbl{eq:grr1}
G_{\rho \rho} (k,z)=\frac{\bar{\rho}}{z}
\left[S(k)
+\frac{G^{(0)}_{\rho{B}}(k,z)}
{1-V(k)G^{(0)}_{\rho{B}}(k,z )}
\right]
\ee

\noindent In the noninteracting limit, we define $\psi(z)$ as the
density autocorrelation function normalized with respect to its equal
time value

\be
\psi (z)=
G^{(0)}_{\rho \rho} (k,z)/\rho_{0}~~.
\ee

\noindent Laplace transforming the FDT relation
(\ref{fdtreln0}) for the noninteracting system we obtain

\be
G^{(0)}_{\rho B}(z)= -\beta {\rho}_0 (1-z\psi (z))
\ee

\noindent Using this we obtain from eqn. (\ref{eq:grr1}) the result

\be
G_{\rho \rho} (k,z) = \frac{\bar{\rho}}{z}
\left[ \frac{1}{1+\beta\rho_0V(k)}
+\frac{(1-z\psi (z))}
{1+\beta \rho_{0}V(k)(1-z\psi (z))}
\right]
~~~.
\ee

\noindent After some algebra we have for the density correlation function,

\be
G_{\rho \rho} (k,z) = \bar{\rho}S(k) \left [ \frac{\psi (z)}
{1+\beta \rho_{0}V(k)\{1-z\psi (z)\}} \right ]
\label{eq:87}
\ee

\noindent where

\be
\psi (z)=-i\int_{0}^{\infty}dt e^{izt}e^{-\frac{1}{2}(kv_{0}t)^{2}}
\ee

\noindent Let us look at this in the small k limit.  We easily find
\be
\psi (z)=\frac{1}{z} \left [ 1+\frac{(kv_{0})^{2}}{z^{2}} \right ]
\ee

\noindent plus terms of order $k^{4}$.  Putting this into Eq.(\ref{eq:87})
and rearranging we find

\be
G_{\rho\rho}(k,z)=\frac{\bar{\rho}S(0)}{2}
\left[\frac{1}{z-ck}
+\frac{1}{z+ck}\right]
\ee

\noindent The two poles  represent  the two propagating sound modes
respectively with speed $c(k)$ is obtained as,

\be
c(k)= v_0 {\{1+\beta \rho_{0}V(k)\}}^{1/2}
=\frac{1}{\sqrt{\beta{m}S(k)}}.
\ee

\section{Momentum Variables, Currents, and the Hierarchial
Structure of Theory}

\subsection{Additional Degrees of Freedom}

So far the theory has been set up to deal with what we will call
the {\it core} problem. This is the determination of the observables
involving the core variables $\Phi_{0}=(\rho , B)$.
In the case of SD this covers essentially all of the degrees
of freedom of interest.  For ND this is not the case. We have
a number of additional degrees of freedom. This is because
we have a larger phase-space due to the momentum degrees of freedom.
Choices of variables to be included from the simplest
to the most involved are:

\noindent 1.  Couple to the two transverse currents $g_{\perp,1}$,
$g_{\perp,2}$

\noindent 2.  Couple to the whole current ${\bf g}$ and/or the
kinetic energy density ${\bf g}_{K}$.

\noindent 3.  Couple to the phase-space density \be
f(x,p,t)=\sum_{i=1}^{N}\delta (p-P_{i}(t))\delta (x-R_{i}(t)~~~. \ee
Choice 1 is the simplest since there is no direct coupling between
the longitudinal and transverse degrees of freedom.  This is also
the simplest way of determining the shear viscosity. If one  goes to
choice 2 things are more complicated since we have all the
correlation functions among $\rho$, $B$ and $g$. One needs ${\bf
g}_{K}$ to determine the thermal conductivity.

If one wants to investigate the Boltzmann equation and momentum
distributions one needs to  include the phase-space density
$f(x,p,t)$ as one of the variables.  This will be discussed elsewhere.

\subsection{Generic Inclusion Of Additional Variables}

Suppose that we have a single-particle additive variable
$g_{\alpha}$ we want to include in our set $\Phi =(\rho , B, g)$.
This inclusion simply involves an additional term in the basic action

\be
A-H\cdot\Phi_{0}\rightarrow
A-H\cdot\Phi_{0}-J\cdot g
\ee

\noindent which introduces a new external coupling $J_{\alpha}$
into the problem.
The fundamental generating functional is now given by
\be
W[H,J]=ln Z_{T}[H,J]
\ee
\noindent where
\be
Z_{T}[H,J]=\sum_{i=1}^{N}\frac{\rho_{0}^{N}}{N!}Z_{N}[H,J]
\ee
\noindent and
\be
Z_{N}[H,J]=Tr^{N}e^{A-H\cdot\Phi_{0}-J\cdot g}
~~~.
\ee
\noindent Functional derivatives with respect to $J$ generates a factor
of $g$ in an average
\be
\langle g_{\alpha}\rangle=\frac{\delta}{\delta J_{\alpha}}
W[H,J]
\ee

\subsection{Perturbation Theory and Hierarchial Structure of Theory}

It is trivial to see that the fundamental identity of the theory
takes the form

\be G_{\alpha}=Tr^{0}\phi_{\alpha}e^{H\cdot \phi_{0}+J\cdot g}
e^{\Delta W[H;J]}
\ee

\noindent where

\be
\Delta W[H;J] =W[H+F;J]-W[H;J]
\ee

\noindent in the extended space.

We can again construct a dynamical OZ equation of the form
\be
G_{\alpha\beta}={\cal G}_{\alpha\beta}+c_{\alpha \mu}G_{\mu \beta}
\ee

\noindent
where the single-particle contribution is given by

\be {\cal G}_{\alpha\beta} =Tr^{0}\phi_{\alpha}\phi_{\beta}e^{H\cdot
\phi_{0}+J\cdot g} e^{\Delta W[H;J]} \ee

\noindent and the dynamic direct correlation function

\be c_{\alpha\mu} =Tr^{0}\phi_{\alpha}e^{H\cdot \phi_{0}+J\cdot g}
e^{\Delta W[H;J]}\frac{\delta}{\delta G_{\mu}}\Delta W[H;J]
\ee

\noindent The key point is that to first order in perturbation
theory

\be \Delta W[H;J]^{(1)}=\sum_{u_{0}}F_{u_{0}}G_{u_{0}}
\ee

\noindent where the sum over $u_{0}$ is only over the core variables
$\phi_{0}= \{\rho , B\}$.
%
%
This happens because the Hamiltonian and hence the corresponding MSR
action involves only these core variables.
The first order dynamical direct correlation function $c_{\alpha
\mu}^{(1)}$ vanishes if $\mu = g$.  This means at first order we
have the kinetic equation

\be G_{\alpha\beta}={\cal G}_{\alpha\beta} +{\cal G}_{\alpha
\rho}\beta V G_{\rho\beta} +{\cal G}_{\alpha B}\beta VG_{B\beta}
~~~.
\label{eq:EPS}
\ee

\noindent The structure of this equation is very interesting.  If we
restrict $\alpha$ and $\beta$ to the core variables then
Eq.(\ref{eq:EPS}) reduces to the first- order core problem we solved
earlier. The correlation functions for the extended variables $\Phi$
can be expressed in terms of (roughly) noninteracting correlation
functions between all the variables and solutions to the core
problem associated with $\Phi_{0}$.

This suggests that one can not get around the core problem.  This
problem is self-contained and  must be treated first.  These
self-consistent solutions then enter into the extended space $\Phi$
as determined sub-matrices in the matrix kinetic equation.
The second order order corrections to the dynamic direct correlation
functions $c^{(2)}_{ij}$ and hence the self energy matrix
$K_{ij}^{(2)}$  will involve the extended set of collective
variables.

\section{Non-interacting system}

\subsection{Generating functional}

In this section we present the calculation of the zeroth- order
cumulants for the
Newtonian dynamics case.
We first work out the generating functional for a rather general
single-particle gaussian model in the Appendix \ref{appB}.  This model includes
SD, ND and Fokker-Planck dynamics as special cases.
For a quadratic action of fields
 $\psi_{i}$ and response fields $\hat{\psi}_{i}$ we have

\bea
A_{0} [\psi,\hat{\psi}] &=& \sum_{ij}\int_{t_{0}}^{\infty}dt
\hat{\psi}_{i}(t)\bar{D}_{ij}\hat{\psi}_{j}(t) +
\sum_{i} \int_{t_{0}}^{\infty}dt
\left[i\hat{\psi}_{i}(t)\left(\dot{\psi}_{i}(t)
+\sum_{j}K_{ij}\psi_{j}(t)\right)\right] \nonumber \\
&-&\sum_{i}\int_{t_{0}}^{\infty}dt
\left[h_{i}(t)\psi_{i}(t)+\hat{h}_{i}(t)\hat{\psi}_{i}(t)\right]
\nonumber \\
\eea

\noindent
where $\bar{D}_{ij}$ is the damping matrix, $K_{ij}$ is a force  matrix,
and $h_{i}$ and $\hat{h}_{i}$ are the detailed external fields
that couple to $\psi_{i}(t)$ and $\hat{\psi}_{i}(t)$.
In Appendix \ref{appB} we find that the associated generating functional
is given by

\bea
\ln Z_{0}(h,\hat{h};\psi^{(0)})
&=& \sum_{ij}\int dt \int dt' \left [
\frac{1}{2} h_{i}(t)c_{ij}(t-t')h_{j}(t')
+h_{i}(t)g_{ij}(t-t')\hat{h}_{j}(t') \right ] \nonumber \\
&+&\sum_{ij}\int dt h_{i}(t)ig_{ij}(t-t_{0})\psi_{j}^{(0)}
\nonumber \\
~~~.
\eea

\noindent where $\psi_{i}^{(0)}$ is the initial value of the fields,
the normalization is such that $Z_{0}[0,0]=1$,

\bea
c_{ij}(t,t') &=& -2\sum_{k,\ell}\int_{-\infty}^{\infty}d\bar{t}g_{ik}(t,\bar{t})
\bar{D}_{k\ell}g_{\ell j}^{T}(\bar{t},t') \nonumber \\
&=& -\sum_{k,\ell}\int_{-\infty}^{\infty}d\bar{t}g_{j\ell}(t',\bar{t})
2\bar{D}_{k\ell}g^T_{\ell{i}}(\bar{t},t')=c_{ji}(t',t)
\eea

\noindent and the function $g_{ij}$ is now obtained from the solution of
the Green's function equation

\be
\lbl{eq:102}
\frac{\partial}{\partial{t}} g_{ij}(t,t')
+\sum_k K_{ik} g_{kj}(t,t')
= -i\delta(t-t')\delta_{ij}
~~~.
\ee

\noindent These equations govern SD, ND and FPD.  Notice that all information
about the equilibrium state of the system is carried by the damping matrix
$\bar{D}\approx k_{B}T$.  If we restrict ourselves to simple fluids,
then the damping term with $c_{ij}$ is zero and the state of the system enters
via the initial condition. We have

\be
Z_{0}(h,\hat{h},\psi^{(0)})=
exp(\int_{t_{0}}^{\infty}dt
\int_{t_{0}}^{\infty}dt'\sum_{ij}
h_{i}(t)g_{ij}(t-t')[\hat{h}_{j})
+i\delta (t'-t_{0})\psi_{j}^{(0)}])
\ee

\noindent with an average over initial conditions remaining.
If the initial conditions are gaussian

\be
P_{0}(\psi^{(0)})={\cal N}e^{-\frac{1}{2}
\psi_{i}^{(0)}M_{ij}\psi_{j}^{(0)}}
\ee

\noindent one can carry out the average over $\psi_{i}^{(0)}$
and obtain

\be
ln Z_{0}(h,\hat{h})=\frac{1}{2}
\int_{t_{0}}^{\infty}dt
\int_{t_{0}}^{\infty}dt'\sum_{ij}
h_{i}(t)c^{T}_{ij}(t-t')h_{j}(t')
+h_{i}(t)g_{ij}(t-t')\hat{h}_{j}
\ee

\noindent and

\bea
c^{T}_{ij}(t,t') &=& c_{ij}(t,t') +c_{ij}^{I}(t,t')
\nonumber \\
c_{ij}^{I}(t,t') &=& \sum_{k\ell}
g_{ik}(t,t_{0})(M^{-1})_{k\ell}g^{T}_{\ell j}(t_{0},t')
~~~.
\eea

\subsection{\bf Newtonian Dynamics}

For Newtonian dynamics there are two phase-space coordinates, $R$ and $P$,
with an elementary force matrix $K_{ij}=\delta_{i,R}\delta_{j,P}\frac{1}{m}$
for the noninteracting system.
We first have to solve the matrix equation following
from Eq.(\ref{eq:102})
in the Newtonian case for which indices $i$ and $j$ run over the
set $\{R,P\}$.

\bea
\lbl{eq:grr}
\frac{\partial}{\partial t}g_{RR}-\frac{g_{PR}}{m} &=& -i\delta (t)\\
\frac{\partial}{\partial t}g_{RP}-\frac{g_{PP}}{m}&=& 0 \nonumber \\
\frac{\partial}{\partial t}g_{PR}&=& 0 \nonumber \\
\frac{\partial}{\partial t}g_{PP}&=& -i\delta (t)
\eea

\noindent The straight forward solution $g_{ij}(t-t')$ is listed
in Table 1.

\begin{center}
\begin{table}
\begin{tabular}{|c|c|c|}
\hline
 & & \\
\ \ \ \ \ \ \ \ \  \ \ \ \ \ & $R$  & $P$\\
\hline
 & & \\
$R$    & $-i\theta (t-t')$  &$-\frac{i}{m}\theta (t-t')(t-t') $ \\
 & & \\
$P$    & $0$ & $-i\theta (t-t')$ \\
 & & \\
\hline
\end{tabular}
\caption[Matrix $g$: Newtonian Case]{Matrix $g$: Newtonian Case.}
\label{tab:1.1}
\end{table}
\end{center}

\noindent Since there is no dissipation put into the model,
$\bar{D}_{ij}=0$, $c_{ij}^{(0)}=0$ the generating functional simplifies

\bea
ln Z_{0}[h,\hat{h};\psi_{i}^{(0)}] &=&
\int_{t_{0}}^{\infty}dt
\int_{t_{0}}^{\infty}dt'\sum_{ij}
h_{i}(t)g_{ij}(t-t')\hat{h}_{j}(t') \nonumber \\
&+& \int_{t_{0}}^{\infty}dt\left[
h_{R}(t)\theta (t-t_{0})R_0
+i\sum_{i} h_{i}(t)ig_{iP}(t-t_{0})P_0 \right]~~.
\eea

\noindent The general result for the solution of the differential equation

\be
\frac{\partial}{\partial{t}} <\psi_i(t)> +\sum_{j} K_{ij} <\psi_j (t)>
= \psi_i^{(0)} \delta(t-t_0)
\ee

\noindent is obtained in terms of the Green's function $g(t,t')$

\be
<\psi_i (t;t_{0})>=\sum_{j} i\int dt' g_{ij} (t-t')\psi_{j}^{(0)}
\delta (t'-t_{0})
= ig_{ij}(t-t_0)\psi_j^{(0)}.
\ee

\noindent
For $c_{ij}=0$ and $\hat{h}=h=0$ we obtain the same relation from
eqn. (\ref{eq:22}).
In this particular case, we obtain the following
reversible ( and deterministic) equations of motion are given for
the two coordinates $\{R,P\}$.
\bea
R(t) &=& ig_{RR}(t-t_{0})R_{0}+ig_{RP}(t-t_{0})P_{0}
\nonumber \\
P(t) &=& ig_{PR}(t-t_{0})R_{0}+ig_{PP}(t-t_{0})P_{0}
\eea

\noindent These equations are also obtained from the relations

\bea
<R(t)> &=& \frac{\delta}{\delta{h_R}}
\ln Z_0 {\Big |}_{h,\hat{h}=0} \nonumber \\
<P(t)> &=& \frac{\delta}{\delta{h_P}}
\ln Z_0 {\Big |}_{h,\hat{h}=0} \nonumber
\eea

\noindent
Inserting the $g$'s leads to the standard phase-space trajectories
for free streaming particles.

\bea
R(t) &=& \theta(t-t_0)
\left [ R_{0}+(t-t_0)\frac{P_0}{m}\right ] \nonumber \\
P(t) &=& \theta(t-t_0)P_0
\eea

\noindent
The generating function for ND is given by

\bea
ln Z_{0}[h,\hat{h}] &=& \int_{t_{0}}^{\infty}dt
\int_{t_{0}}^{\infty}dt'\sum_{ij}h_{i}(t)g_{ij}(tt')\hat{h}_{j}(t')
\nonumber \\
&+& \int_{t_{0}}^{\infty}dt(h_{R}(t)\theta (t-t_{0})R_0
+\int_{t_{0}}^{\infty}dt \sum_{i}h_{i}(t)g_{iP} (t-t_{0})P_0
~~~.
\eea

\subsection{Computation of $\phi$-Correlations}

It is not the direct correlations of the phase-space coordinates
that are of interest but the correlations of the $\Phi$.  It
is clear that we need to evaluate all of the $\rho -B$
cumulants generated by

\be
W_{0}[H]=Tr e^{H\cdot \phi }
~~~.
\ee

\noindent
We begin by introducing the  microscopic
sources $h_{i}$, $\hat{h}_{i}$ and treating

\be
Z_{0}[H,h,\hat{h}]=Tr e^{H\cdot \phi }
e^{h\cdot\psi +\hat{h}\cdot\hat{\psi}}
~~~.
\ee

\noindent Remember that the spatial Fourier transforms of the $\phi$ are
in terms of the phase-space coordinates $\psi$:

\bea
\phi_{\rho} (1) &=& e^{-ik_{1}R(t_{1})} \nonumber \\
\phi_{B} (1) &=& -k_{1}\cdot \hat{P}(t_1) e^{-ik_{1}.R(t_{1})}~~~.
\eea

\noindent Introducing the operators

\bea
\hat{\phi}_{\rho}(1) &=& e^{-ik_{1}\frac{\delta}{\delta h_{R}(t_1)}}  \\
\hat{\phi}_{B}(1) &=& \hat{b}(1)\hat{\phi}_{\rho}(1) \noindent \\
\eea

\noindent where

\be
\hat{b}(1)=-k_{1}\cdot \frac{\delta}{\delta \hat{h}_P(t_1)}
~~.
\ee

\noindent We can write

\be
Z_{0}[H,h,\hat{h}]=Tr e^{H\cdot\phi+h\cdot\psi+\hat{h}\cdot\hat{\psi}}
\nonumber
\ee
\be
=e^{H\cdot\hat{\phi}}
Tr e^{h\cdot\psi+\hat{h}\cdot\hat{\psi}}
\nonumber
\ee
\be
=e^{H\cdot\hat{\phi}}
\rho_{0}e^{h\cdot g\cdot\hat{h}}
\int d\psi_{0}P[\psi_{0}]e^{h\cdot ig \psi_{0}}
\ee
\be
=e^{H\cdot\hat{\phi}}
\rho_{0}
\int d\psi_{0}P[\psi_{0}]e^{h\cdot g\cdot [\hat{h}+ i\psi_{0}]}
\ee

One can then express an arbitrary noninteracting
cumulant, in the presence of $h$
and $\hat{h}$, in the form

\be
G_{BB\ldots B\rho\ldots\rho}(12\ldots \ell \ell +1\ldots n;h,\hat{h})
=\hat{b}(1)\ldots \hat{b}(\ell )\hat{\phi}(1)\ldots \hat{\phi}(n)
Z_{0}[h,\hat{h}]
\nonumber
\ee
\be
=\hat{b}(1)\ldots \hat{b}(\ell )\hat{\phi}(1)\ldots \hat{\phi}(n)
\rho_{0}
\int d\psi_{0}P[\psi_{0}]e^{h\cdot g\cdot [\hat{h}+ i\psi_{0}]}
\ee

\noindent The analysis is very similar to that for the SD case in Appendix C in
Ref. \cite{FTSPD}.  The $\hat{\phi}_{\rho}$ are translation operators:

\be
G_{BB\ldots B\rho\ldots\rho}(12\ldots \ell \ell +1\ldots n;h,\hat{h})
=\hat{b}(1)\ldots \hat{b}(\ell )
\rho_{0}
\int d\psi_{0}P[\psi_{0}]e^{[h+L]\cdot g\cdot [\hat{h}+ i\psi_{0}]}
\ee

\noindent where in more detail

\be
 h+L\rightarrow h_{\alpha}(t_{j})+L_{\alpha}^{(n)}(t_{j})
\ee

\noindent with

\be
 L_{\alpha}^{(n)}(t_{j})=-i\delta_{\alpha ,R}
\sum_{s=1}^{n}k_{s}\delta (t_{j}-t_{s})
\ee

\noindent Application of the operators $\hat{b}(i)$ is multiplicative

\be
G_{BB\ldots B\rho\ldots\rho}(12\ldots \ell \ell +1\ldots n;h,\hat{h})
=b_{n}(1)\ldots b_{n}(\ell )
\rho_{0}
\int d\psi_{0}P[\psi_{0}]e^{[h+L]\cdot g\cdot [\hat{h}+ i\psi_{0}]}
\ee

\noindent and

\be
b_{n}(j)=-k_{j}\int_{t_{0}}^{\infty}d\bar{t}
\sum_{\alpha}\left[h_{\alpha}(\bar{t}+L_{\alpha}^{n}(\bar{t})\right]
g_{\alpha P}(\bar{t},t_{j})
~~~.
\ee

\noindent Thus we have the cumulants for arbitrary $h$ and $\hat{h}$.
Setting $h$ and $\hat{h}$ to zero we have

\be
G_{BB\ldots B\rho\ldots\rho}(12\ldots \ell \ell +1\ldots n)
=b_{n}(1)\ldots b_{n}(\ell )
~~~.
\rho_{0}
\int d\psi_{0}P[\psi_{0}]e^{L\cdot g\cdot i\psi_{0}]}
\ee
\be
=b_{n}(1)\ldots b_{n}(\ell )
\rho_{0}
\int d\psi_{0}P[\psi_{0}]e^{\kappa_{\alpha}\psi_{0}^{\alpha}}
\ee

\noindent where

\be
\kappa_{\alpha}=iL\cdot g
=iL_{\mu}^{(n )}(\bar{t})g_{\mu\alpha}(\bar{t},t_{0})
\ee

\noindent and

\be
b_{n}(j)=-k_{j}\int_{t_{0}}^{\infty}d\bar{t}
(-i\sum_{s=1}^{n}k_{s}\delta (\bar{t}-t_{s}))
\left(-\frac{i}{m}\theta (\bar{t}-t_{j})(\bar{t}-t_{j})\right)
\nonumber
\ee
\be
=\frac{k_{j}}{m}
\sum_{s=1}^{n}k_{s}
\theta (t_{s}-t_{j})(t_{s}-t_{j})
\nonumber
\ee

\noindent We have after a little work

\be
\kappa_{R}=-i
\sum_{s=1}^{n}k_{s}
\ee
\be
\kappa_{P}=-i
\sum_{s=1}^{n}k_{s}(t_{s}-t_{0})
\nonumber
\ee
\be
=-i T
\ee
and
\be
G_{BB\ldots B\rho\ldots\rho}(12\ldots \ell \ell +1\ldots n)
=b_{n}(1)\ldots b_{n}(\ell )
\nonumber
\ee
\be
\times\rho_{0}
\int d^{d}R_{0}P[R_{0}]e^{i\sum_{s=1}^{n}k_{s}\cdot R_{0}}
\int d^{d}P_{0}P[P_{0}]e^{-iT\cdot P_{0}}
\ee
\be
=b_{n}(1)\ldots b_{n}(\ell )
\rho_{0}
(2\pi )^{d}\left(\sum_{s=1}^{n}k_{s}\right)
e^{N_{n}}
\ee
where
\be
e^{N_{n}}=\int d^{d}P_{0}P[P_{0}]e^{-iT\cdot P_{0}}
~~~.
\ee

\noindent The probability of finding a particle at a given position
is uniform and the position
average enforces the translational invariance in the system

\be
\int d^{d}R_{0}e^{\kappa_{R}\cdot R_{0}}=
 (2\pi)^{d}\delta  (k_{1}+ k_{2}+\ldots+ k_{n})
~.
\label{eq:134}
\ee

\noindent The momentum is assumed to obey a Maxwell-Boltzmann distribution and
the average gives the contribution
\be
e^{N_{n}}=\int\frac{d^{d}P}{(2\pi p_{0})^{d/2}}
e^{-\beta\frac{P^{2}}{2m}}e^{-iT\cdot P}
\ee

\noindent where $p_{0}^{2}=m \beta^{-1}=m^2v_0^2$ is the thermal
momentum.
This is a complete the square calculation giving

\be
e^{N_{n}}= e^{-\frac{1}{2}p_{0}^{2}T^{2}}
\ee

\noindent and we identify the effective $N_{n}$

\be
N_{n}=-\frac{1}{2}p_{0}^{2}T^{2}
~~~.
 \ee

\noindent The correlation function $G_{BB\ldots B\rho\ldots\rho}$
with $\ell$ number of
$B$ fields is obtained as

\bea
G_{BB\ldots B\rho\ldots\rho}(12\ldots \ell \ell +1\ldots n)
&=& b_{n}(1)\ldots b_{n}(\ell )
\rho_{0}
(2\pi )^{d}\left(\sum_{s=1}^{n}k_{s}\right)
e^{N_{n}}
~~~.
\eea

\noindent Let us focus on $N_{n}$.  Writing it out we have
\be
N_{n}=-\frac{p_{0}^{2}}{2m^2}
\left[\sum_{s=1}^{n}k_{s}(t_{s}-t_{0})\right]^{2}~~.
 \ee

\noindent Using the conservation law Eq. (\ref{eq:134}) we see this
expression is independent
of $t_{0}$ and is time translationally invariant. The effective $N_{n}$
can be written in a form similar to the SD case.
We can rewrite $N_{n}$

\be
N_{n}=-\frac{1}{2}\sum_{i=1}^{n}\sum_{j=1}^{n}k_{i}k_{j}C_{ij}
 \ee

\noindent where

\be
C_{ij}=\frac{p_0^2}{m^2}t_{i}t_{j}
\ee

\noindent Then define

\be
D_{ij}=C_{ii}+C_{jj}-2C_{ij}
\ee
\be
=v_0^2(t_{i}-t_{j})^{2}
~~~.
\ee

\noindent We then have

\bea
N_{n} &=& -\frac{v_{0}^{2}}{2}\left[\sum_{i=1}^{n}k_{i}^{2}C_{ii}
+\sum_{i\neq j =1}^{n}\sum_{j=1}^{n}k_{i}k_{j}C_{ij}\right]
\nonumber \\
&=& -\frac{v_{0}^{2}}{2}\left[\sum_{i=1}^{n}k_{i}^{2}C_{ii}
+\sum_{i\neq j =1}^{n}k_{i}k_{j}\frac{1}{2}
[C_{ii}+C_{jj}-D_{ij}]\right] \nonumber \\
&=& -\frac{v_{0}^{2}}{2}\left[\sum_{i=1}^{n}k_{i}^{2}C_{ii}
+\sum_{i\neq j =1}^{n}k_{i}k_{j}\frac{1}{2}C_{ii}
+\sum_{i\neq j =1}^{n}k_{i}k_{j}\frac{1}{2}[C_{jj}
+\sum_{i\neq j =1}^{n}k_{i}k_{j}\frac{1}{2}(-D_{ij})\right] \nonumber \\
&=& -\frac{v_{0}^{2}}{2}\left[\sum_{i=1}^{n}k_{i}^{2}C_{ii}
-2\sum_{i=1}^{n}k_{i}^{2}\frac{1}{2}C_{ii}
+\sum_{i\neq j =1}^{n}k_{i}k_{j}\frac{1}{2}(-D_{ij})\right]
\nonumber \\
&=& \frac{v_{0}^{2}}{4}\sum_{i=1}^{n}\sum_{j=1}^{n}
K_{ij}(t_{i}-t_{j})^{2}
\eea
where
\be
K_{ij}=k_{i}\cdot k_{j}
~~~.
\ee
The noninteracting correlation and response cumulants are now
all in a time-translationally invariant form.

\subsection{Two-Point Cumulants}

Let us extract the two-point cumulants needed in our first-order
discussion.  For the density-density correlation function
\be
G_{\rho\rho}^{(0)}(12)=\rho_{0}(2\pi )^{d}\delta (k_{1}+k_{2})
e^{-\frac{1}{2}\kappa_{1}v_{0}^{2}(t_{1}-t_{2})^{2}}
\ee
\noindent where
\be
\kappa_{1}=k_{1}^{2}
~~~.
\ee
\noindent For the response function we have
\be
G_{\rho B}^{(0)}(12)=b_{2}(2)G^{(0)}_{\rho\rho}(12)
\ee
\noindent where
\be
b_{2}(2)=-\frac{\kappa_{1}}{m}\theta (t_{1}-t_{2})(t_{1}-t_{2})
\ee
\noindent and
\be
G_{BB}^{(0)}(12)=b_{2}(1)b_{2}(2)G^{(0)}_{\rho\rho}(12)
=0
~~~.
\ee
\noindent Notice the relationship
\be
G_{\rho B}^{(0)}(12)=\theta (t_{1}-t_{2})\beta
\frac{\partial}{\partial t_{1}}
G^{(0)}_{\rho\rho}(12)
~~~.
\ee
\noindent holds at the zeroth order. We return to these results in the next paper in this series.

\section{Conclusions}

We have developed the fundamental theory for conventional
fluids in which the particles follow reversible Newtonian
dynamics. The theory is remarkably similar to
that developed to treat particles following Smoluchowski dynamics(SD).
We have a field theoretic formulation with similar nonlinear
interactions in terms of a pair potential. The question of
self-consistency is addressed
in a very similar manner.  The role of initial conditions
and broken time translational symmetry are addressed at zeroth
and first order in the theory.  At zeroth order the average over initial
conditions is shown for cumulants of the collective
variables to satisfy TTI. At first order we can impose equilibrium
and sustain TTI if we require the system to obey fluctuation
dissipation symmetry (FDS).

As for the treatment of SD the second order self-energies require
the zeroth order irreducible three-point vertices. These
three-point functions turn out to be more complicated in the case of
ND. In the present paper we have shown how to compute all cumulants
for the non interacting system.
The three-point cumulants enter into the determination of the
three-point irreducible vertices.  These vertices enter into
the computation of the second order self-energy.
 This becomes the necessary input for analysis of a possible ENE
 transition in such systems at high density.
In the next paper in this series we show that there is a FDS and
it can rather easily be applied to the cumulants and irreducible
vertices
of the fully interactive fluid ND system.  We derive identities
obeyed by the three-point objects in a manner similar to the
application of nonperturbative FDS to the two-point cumulants and vertices.
We then focus on the use of this machinery to explore
whether we have ENE transitions in ND systems.

\section*{Acknowledgements} We thank the Joint Theory Institute  at the
University of Chicago for their generous support.  We also thank
David Mc Cowen and Paul Spyridis for comments and help.

\newpage
\appendix

\section{Fundamental Identities}
\label{appA}

We have the definition of the grand partition function
\be
Z_{T}[H;x]=\sum_{N=0}^{\infty}\frac{x^{N}}{N!}Tr^{N}
e^{H\cdot\Phi_{N}+\frac{1}{2}\Phi_{N}\sigma\Phi_{N}}
\ee
\noindent where
$\Phi_{N}=\sum_{i=1}^{N}\phi^{(i)}$
and we have the important result that self-energies ( the interaction of the
$i$-th particle with itself does not contribute, {\em i.e.},
\be
\phi^{(i)}\sigma\phi^{(i)}=0
\ee
\noindent for all $i$. We need the following identity

\be
\Phi_{N+1}\sigma\Phi_{N+1}
=\Phi_{N}\sigma\Phi_{N}+2\phi^{(N+1)}\sigma\Phi_{N}
=\Phi_{N}\sigma\Phi_{N}+2F^{N+1}\Phi_{N}
\ee
\noindent where we have defined $F^{(N+1)}=\sigma \phi^{(N+1)}$.
Therefore we have

\be
\frac{1}{2}\Phi_{N+1}\sigma\Phi_{N+1}
=\frac{1}{2}\Phi_{N}\sigma\Phi_{N}
+F^{(N+1)}\Phi_{N}
~~~.
\ee
\noindent Now consider
\be
Z_{N}(H+F_{N+1})=
Tr^{N}e^{(H+F^{(N+1)})\cdot\Phi_{N}+\frac{1}{2}\Phi_{N}\sigma\Phi_{N}}
\nonumber
\ee
\be
=Tr^{N}e^{H\cdot\Phi_{N}+\frac{1}{2}\Phi_{N+1}\sigma\Phi_{N+1}}
~~~.
\ee
\noindent Multiplying by $e^{H\cdot \phi_{N+1}}$ gives the result
\be
e^{H\cdot \phi_{N+1}}Z_{N}(H+F^{(N+1)})
=Tr^{N}e^{H\cdot\Phi_{N+1}+\frac{1}{2}\Phi_{N+1}\sigma\Phi_{N+1}}
~~~.
\ee
\noindent This is the canonical form of the central identity. We now
Trace over the degrees of freedom of the $(N+1)$-th particle
( which we denote by the label $0$, {\em i.e.}, $N+1\rightarrow 0$)

\bea
\tilde{T}r^{(0)}e^{H\cdot \phi^{(0)}}Z_{N}(H+F^{(0)})
&=& Tr^{N+1}e^{H\cdot\Phi_{N+1}+\frac{1}{2}\Phi_{N+1}\sigma\Phi_{N+1}}
\nonumber \\
&\equiv& Z_{N+1}(H)
\eea

\noindent where $\tilde{T}r^{(0)}$ implies the trace taken over the phase
 space coordinates of the $'0'$-th particle.
 We now multiply the above equation by $\frac{x^{N}}{N!}$
and sum over all $N$

\bea
\lbl{eq:zT1}
&&\tilde{T}r^{(0)} e^{H\cdot \phi^{(0)}}Z_{T}(H+F^{(0)})
= \sum_{N=0}^{\infty}\frac{x^{N}}{N!}Z_{N+1}(H)
= \sum_{N=1}^{\infty}\frac{x^{N-1}}{(N-1)!}Z_{N}(H) \nonumber \\
&=& \frac{\partial}{\partial x}
\sum_{N=0}^{\infty}\frac{x^{N}}{N!}Z_{N}(H)
= \frac{\partial}{\partial x}Z_{T}[H;x] ~~.
\eea

\noindent Divide the above equation by $Z_{T}[H;x]$ to obtain

\be
\lbl{eq:zT11}
\frac{\partial}{\partial x}W [H;x]
=\tilde{T}r^{(0)}e^{H\cdot \phi^{(0)}+\Delta W[H+F^{(0)};x]}
\ee

\noindent where $\ln Z_T [H;x]=W[H;x]$ and we have defined

\be
\Delta W[H+F;x]=W[H+F^{(0)};x]-W[H;x]~~.
\ee

\noindent Integrating the relation (\ref{eq:zT1}) we obtain

\be
W(H;\rho_{0})=\int_{0}^{\rho_{0}}dx \tilde{T}r^{(0)}
e^{H\cdot\phi^{(0)}+\Delta W[H+F^{(0)};x]}
~~~.
\ee
\noindent We use this result to generate the one-point quantity

\bea
G_{\alpha} &=& \frac{\delta W}{\delta H_{\alpha}}
=\frac{\delta }{\delta H_{\alpha}}
\int_{0}^{\rho_{0}}dx \tilde{T}r^{(0)}
e^{H\cdot\phi^{(0)}+\Delta W(H+F^{(0)};x)}
\nonumber \\
&=& \int_{0}^{\rho_{0}}dx \tilde{T}r^{(0)} \phi^{(0)}_{\alpha}
e^{H\cdot\phi^{(0)}+\Delta W(H+F^{(0)};x)}
\nonumber \\
&+& \int_{0}^{\rho_{0}}dx \tilde{T}r^{(0)}
e^{H\cdot\phi^{(0)}+\Delta W(H+F^{(0)};x)}
\left[G_{\alpha}(H+F^{(0)};x)-G_{\alpha}(H;x)\right]~~.
\label{galpha1}
\eea

\noindent Substituting $H'=H+F^{(0)}$ the integral in the second term on the
RHS is written as

\be
\int_{0}^{\rho_{0}}dx \tilde{T}r^{(0)}
e^{H\cdot\phi^{(0)}+\Delta W(H';x)}
\left[G_{\alpha}(H';x)-G_{\alpha}(H;x)\right]~~~.
\ee

\noindent  Consider first the first part as,

\bea
I_{2} &=& \int_{0}^{\rho_{0}}dx \tilde{T}r^{(0)}
e^{H\cdot\phi^{(0)}}e^{\Delta W(H';x)}
G_{\alpha}(H';x) \nonumber \\
&=& \int_{0}^{\rho_{0}}dx
 \tilde{T}r^{(0)}e^{H\cdot\phi^{0}}
\frac{Z_{T}(H';x)}{Z_{T}(H;x)}G_{\alpha}(H';x)
\lbl{galpha2}
\eea

\noindent  We want  to show self-consistently that

\bea
\lbl{aa1}
G_{\alpha}(H';x) &=& Tr^{(1)}\phi_{\alpha}^{(1)}e^{H'\cdot\phi^{(1)}
+\Delta W[H'+F^{(1)}];x} \nonumber \\
&\equiv& x\tilde{T}r^{(1)}\phi_{\alpha}^{(1)}e^{H'\cdot\phi^{(1)}
+\Delta W[H'+F^{(1)}]}
\lbl{galpha3}
\eea

\noindent satisfies the eqn. (\ref{galpha1}) above. Substituting eqn. (\ref{galpha3})
in eqn. (\ref{galpha2}) we obtain

\bea
I_{2}&=& x\tilde{T}r^{(0)}e^{H\cdot \phi^{(0)}}
\frac{Z_{T}(H')}{Z_{T}(H)}
\tilde{T}r^{(1)}\phi_{\alpha}^{(1)}e^{H'\cdot \phi^{(1)}}
\frac{Z_{T}(H'+F^{(1)})}{Z_{T}(H')} \nonumber \\
&=& x\tilde{T}r^{(1)}\frac{\phi_{\alpha}^{(1)}}{Z_{T}(H)}
e^{H'\cdot \phi^{(1)}} \tilde{T}r^{(0)}e^{H\cdot \phi^{(0)}}
Z_{T}(H+F^{(0)}+F^{(1)})\nonumber \\
&=& x\tilde{T}r^{(1)}\frac{\phi_{\alpha}^{(1)}}{Z_{T}(H)}
e^{H\cdot \phi^{(1)}} \tilde{T}r^{(0)}e^{(H+F^{(1)})\cdot \phi^{(0)}}
Z_{T}(H+F^{(1)}+F^{(0)})
\lbl{galpha4}
\eea

\noindent where in getting the last equality we have used the result

\bea
H'\cdot \phi^{(1)}&=& H\cdot\phi^{(1)}+ F^{(0)}\phi^{(1)}
\nonumber \\
&=& H\cdot\phi^{(1)}+ \sigma \phi^{(0)}\phi^{(1)}
\nonumber \\
&=& H\cdot\phi^{(1)}+ F^{(1)}\phi^{(0)} \nonumber
\eea

\noindent Using the last relation (\ref{eq:zT1}), we obtain from eqn. (\ref{galpha4})

\be
I_{2}= x\tilde{T}r^{(1)}\frac{\phi_{\alpha}^{(1)}}{Z_{T}(H)}
e^{H\cdot \phi^{(1)}}\frac{\partial}{\partial x}Z_{T}[H+F^{(1)}]
~~~.
\ee

\noindent
The expression for $I_2$  is put back into the expression (\ref{galpha1})
for $G_{\alpha}$.

\bea
G_{\alpha} &=& \int_{0}^{\rho_{0}}dx \tilde{T}r^{(0)}e^{H\cdot \phi^{(0)}}
\phi_{\alpha}^{(0)}\frac{Z_{T}[H+F^{(0)}]}{Z_{T}[H;x]}
\nonumber \\
&+& \int_{0}^{\rho_{0}}dx x
\tilde{T}r^{(1)}\phi_{\alpha}^{(1)} e^{H\cdot\phi^{(1)}}
\Bigg [ \frac{1}{Z_{T}[H;x]}\frac{\partial}{\partial x}Z_{T}[H+F^{(1)}]
-\frac{Z_{T}[H+F^{(1)}]}{Z^2_{T}[H;x]}\frac{\partial}{\partial x}Z_{T}[H]
\Bigg ] \nonumber \\
&=&\tilde{T}r^{(1)}\phi_{\alpha}^{(1)}
e^{H\cdot\phi^{(1)}}
\int_{0}^{\rho_{0}}dx \Biggl[\frac{Z_{T}[H+F^{(1)}]}{Z_{T}[H]}
+ x\frac{\partial}{\partial x}\frac{Z_{T}[H+F^{(1)}]}{Z_{T}[H;x]}
\Biggr ] \nonumber \\
&=& \tilde{T}r^{(1)}\phi_{\alpha}^{(1)}
e^{H\cdot\phi^{(1)}}
\int_{0}^{\rho_{0}}dx
\frac{\partial}{\partial x}
\Biggl[x \frac{Z_{T}[H+F^{(1)};x]}{Z_{T}[H;x]}\Bigg ]
\nonumber \\
&=& \tilde{T}r^{(1)}\phi_{\alpha}^{(1)}
\rho_{0}e^{H\cdot\phi^{(1)}}
\frac{Z_{T}[H+F^{(1)};\rho_0]}{Z_{T}[H;\rho_{0}]}
\nonumber \\
&\equiv& Tr^{(1)}\phi_{\alpha}^{(1)}
e^{H\cdot\phi^{(1)}+\Delta W[H+ F^{(1)}]}
~~~.
\eea

\noindent Thus we have established the fundamental result
\be
G_{\alpha}=Tr^{(1)}
\phi_{\alpha}^{(1)}
e^{H\cdot\phi^{(1)}+\Delta W[H+ F^{(1)}]}
~~~.
\ee

\section{Single-particle Gaussian Problem}
\label{appB}

We want to construct the generating functional associated
with  single-particle noninteracting dynamics.
The single-particle problem is
governed by the phase-space coordinates $\psi_{i}$
with response variables $\hat{\psi}_{i}$. The action governing
these variables is quadratic

\bea
A_{0} &=& \sum_{ij}\int_{t_{0}}^{\infty}dt
\hat{\psi}_{i}(t)\bar{D}_{ij}\hat{\psi}_{j}(t) +
\sum_{i} \int_{t_{0}}^{\infty}dt
\left[i\hat{\psi}_{i}(t)\left(\dot{\psi}_{i}(t)
+\sum_{j}K_{ij}\psi_{j}(t)\right)\right] \nonumber \\
&-&\sum_{i}\int_{t_{0}}^{\infty}dt
\left[h_{i}(t)\psi_{i}(t)+\hat{h}_{i}(t)\hat{\psi}_{i}(t)\right]
\nonumber \\
\eea

\noindent where $\bar{D}_{ij}$ is the damping matrix, $K_{ij}$ is a force  matrix,
and $h_{i}(t)$ and $\hat{h}_{i}(t)$ are the detailed external fields
that couple to particle $\psi$ and $\hat{\psi}$.
Newtonian dynamics corresponds to the special case where
the damping matrix vanishes $\bar{D}_{ij}=0$.

We  proceed using the  identities that hold  in the range $t_{0}< t < \infty$:

\bea
\int {\cal D}(\psi){\cal D}(\hat{\psi})\frac{\delta}{\delta \psi_{i}(t)}
e^{-A_{0}} &=& 0 \nonumber \\
\int {\cal D}(\psi){\cal D}(\hat{\psi})\frac{\delta}{\delta \hat{\psi}_{i}(t)}
 e^{-A_{0}}&=& 0 \nonumber
\eea

\noindent which leads to the set of equations where we sum over repeated
indices labeled by j and we suppress the local time label:

\bea
2\bar{D}_{ij}\hat{G}_{j}
+i\dot{G}_{i}+iK_{ij}G_{j} &=& \hat{h}_{i}
\label{eq:4} \nonumber \\
-i\frac{\partial}{\partial t}\hat{G}_{i}
+iK^T_{ij}\hat{G}_{j} &=& h_{i}
\label{eq:5a}
 \nonumber
\eea

\noindent where $K^T$ is the transpose of the matrix $K$. We have defined the
functions $G_i$ and $\hat{G}_i$ as

\bea
G_{i} &=& \langle \psi_{i}\rangle
\nonumber \\
\hat{G}_{i} &=& \langle \hat{\psi}_{i}\rangle \nonumber
~~~.
\eea

\noindent The $G$s depend on initial data which we must eventually average over.
We must now solve these equations, at least formally, to obtain the
generating functional.

Let us treat $G_{i}(t)$ and $\hat{G}_{i}(t)$ as inner products spanned by
a complete and orthonormal set of states $|i>$ such that

\bea
G_{i}(t) &=& <i|G(t)> \nonumber \\
\hat{G}_{i}(t) &=& <i|\hat{G}(t)> \nonumber \\
K_{ij} &=& <i|K|j> \nonumber \\
\bar{D}_{ij} &=& <i|\bar{D}|j> \nonumber \\
h_{i}(t) &=& <i|h(t)> \nonumber \\
\hat{h}_{i}(t) &=& <i|\hat{h}(t)> \nonumber
\eea

\noindent which introduces the operators and vectors $G, \hat{G}, K, \bar{D}, h, \hat{h}$.
We then have the operator equations

\bea
\lbl{eq:G}
2\bar{D}\hat{G}(t) +i\left(\frac{\partial}{\partial t}+K\right)G(t)
&=& \hat{h}(t) \\
\lbl{eq:Ghat}
-i\left(\frac{\partial}{\partial t}+K^{T}\right)\hat{G}(t)
&=& h(t) ~~.
\eea

\noindent We first solve the eqn. \ref{eq:Ghat} for $\hat{G}$, obtaining

\be
\hat{G}(t)=\int_{-\infty}^{\infty}d\tau g^{T}(t,\tau )h(\tau )
\ee

\noindent where

\be
g^{T}(t,t')=-ie^{-K^{T} t}\theta (t'-t)e^{K^{T} t'}
~~~.
\ee

\noindent The solution for $\hat{G}$ is put back in the eqn.
(\ref{eq:G}) for $G$ to obtain

\be
\lbl{eq:G1}
\left [ \frac{\partial}{\partial t}+K\right ] \hat{G}(t)
= -i \left ( \hat{h}(t)-2\bar{D}\int_{-\infty}^{\infty}dt' g^T(t,t')h(t') \right ).
\ee

\noindent  It is straightforward to obtain the corresponding
solution for $G(t)$ as

\be
G(t)=i g(t,t_{0})\psi^{(0)}+\int_{-\infty}^{\infty}dt' [ g(t,t')\hat{h}(t')+c(t,t')h(t') ]
\ee

\noindent where the functions $g$ and $c$ are given by

\bea
\lbl{eq:gtt}
g(t,t') &=& -ie^{-Kt}\theta (t-t')e^{Kt'} \\
\lbl{eq:ctt}
c(t,t') &=& -\int_{-\infty}^{\infty}d\bar{t}g(t,\bar{t})2\bar{D}g^{T}(\bar{t},t')
\eea

\noindent
and $\psi^{(0)}$ is the initial value of the phase-space
coordinates.
Notice that $g$ satisfies the Green's function equation

\be
\lbl{gdif-eqn}
\Big [ \frac{\partial}{\partial t}+K  \Big ] g(t,t')=-i\delta (t-t')
\ee

\noindent Putting in complete sets of states we obtain

\bea
\hat{G}_{i}(t ) &=& \sum_{j}\int dt'h_{j}(t' )
g_{ji}(t'-t)
\label{eq:21} \\
G_{i}(t) &=& \sum_j \int_{-\infty}^{\infty}dt' [ g_{ij}(t-t' )\hat{h}_{j}(t' )
+c_{ij}(t-t' )h_{j}(t')]
+ig_{ij}(t-t_{0} )\psi_{j}^{(0)}
\label{eq:22}
\eea

\noindent where the kernel matrix $c_{ij}$ is obtained as

\bea
c_{ij}(t,t') &=& -\sum_{k,\ell}\int_{-\infty}^{\infty}d\bar{t}g_{ik}(t,\bar{t})
2\bar{D}_{k\ell}g_{\ell j}^{T}(\bar{t},t') \nonumber \\
&=& -\sum_{k,\ell}\int_{-\infty}^{\infty}d\bar{t}g_{j\ell}(t',\bar{t})
2\bar{D}_{k\ell}g^T_{\ell{i}}(\bar{t},t')=c_{ji}(t',t)
~~~.
\eea

\noindent In getting the last equality we have used the symmetry
$\bar{D}_{ij}=\bar{D}_{ji}$ of the damping matrix. The function $g_{ij}$ is
now obtained from the solution of
of eqn. (\ref{gdif-eqn}) with labels restored:

\be
\frac{\partial}{\partial{t}} g_{ij}(t,t')
+\sum_k K_{ik} g_{kj}(t,t')
= -i\delta(t-t')\delta_{ij}
\ee

\noindent We then have the results for the generating functional
\bea
\label{eq:23a}
\hat{G}_{i}(t ) &=& \frac{\delta }{\delta \hat{h}(t)}
\ln Z_{0}(h,\hat{h})
\label{eq:23b} \\
G_{i}(t )&=& \frac{\delta }{\delta h(t)}\ln Z_{0}(h,\hat{h})
\label{eqn:24}
\eea

\noindent The solution to this set of equations, Eqs.(\ref{eq:21}),
(\ref{eq:22}), (\ref{eq:23}) and(\ref{eqn:24})
for the generating functional is given by

\bea
\label{eq:23c}
\ln Z_{0}(h,\hat{h};\psi^{(0)})
&=&\frac{1}{2}\sum_{ij}\int dt \int dt' \left [
\frac{1}{2} h_{i}(t)c_{ij}(t-t')h_{j}(t')
+h_{i}(t)g_{ij}(t-t')\hat{h}_{j}(t') \right ] \nonumber \\
&+&\sum_{ij}\int dt h_{i}(t)ig_{ij}(t-t_{0})\psi_{j}^{(0)}
\nonumber \\
&\equiv& \frac{1}{2} h\cdot c \cdot h +h\cdot g\cdot \hat{h}
+h\cdot ig\cdot\psi^{(0)}
~~~.
\eea

\noindent
The last short-hand representation is useful.
The full generator requires averaging over the initial conditions

\bea
Z_{0}[h,\hat{h}] &=& \int d^{d}\psi^{(0)}P_{0}(\psi^{(0)})
\exp \left [ \frac{1}{2}h\cdot c \cdot h +h\cdot g\cdot \hat{h}
+h\cdot ig\cdot\psi^{(0)} \right ]
\nonumber \\
&=& e^{\frac{1}{2}h\cdot c \cdot h +h\cdot g\cdot \hat{h}}
\int d\psi^{(0)}P_{0}[\psi^{(0)}]e^{h\cdot ig\cdot\psi^{(0)}}
~~~.
\eea

\noindent This is the solution to a rather general gaussian problem.


\begin{thebibliography}{99}
\bibitem{close}  If $G_{\rho\rho}$ is the density autocorrelation
function, and we define the limit
\be
\lim_{t\rightarrow\infty}G_{\rho\rho}(q, t)= F(q)
\ee
then if the system is ergodic and $F(q)=0$, while if the system
is nonergodic $F(q)$ is nonzero.
\bibitem{17} The actual definitions of the variables $B$ are different
in the two cases SD and ND.
\bibitem{FDS}  The fluctuation dissipation theorem gives
relationships between cumulants with response fields and those
without. We refer to a fluctuation-dissipation symmetry
(FDS) to characterize the transformation which leaves the action $A$
invariant.  See SDENE for the SD case.
\end{thebibliography}
\end{document}